\documentclass[fleqn,usenatbib]{mnras}

\usepackage{newtxtext,newtxmath}

\usepackage[T1]{fontenc}
\usepackage{balance}
\usepackage{float}
\usepackage{tablefootnote}
\usepackage{subfigure}
\usepackage{caption}

\usepackage[dvipsnames]{xcolor}

\usepackage{flushend}

\newcommand{\beq}[1]{\begin{equation}\label{#1}}
\newcommand{\eeq}{\end{equation}}
\newcommand{\beqn}[1]{\begin{eqnarray}\label{#1}}
\newcommand{\eeqn}{\end{eqnarray}}
\newcommand{\sub}[1]{_\mathrm{#1}}

\newcommand{\tage}{t\sub{age, \star}}
\newcommand{\Op}{\Omega\sub{p}}
\newcommand{\Os}{\Omega\sub{\star}}
\newcommand{\Obar}{\Bar{\Omega}\sub{\star}}
\newcommand{\Ocri}{\Omega\sub{critical}}
\newcommand{\Rc}{R\sub{c}}
\newcommand{\Mc}{M\sub{c}}

\newcommand{\gcons}{\mathcal{G}}

\newcommand{\Osini}{\Omega\sub{ini, \star}}

\newcommand{\Rp}{R\sub{p}}

\newcommand{\Mp}{M\sub{p}}

\newcommand{\Mprate}{\dot{M}\sub{p}}

\newcommand{\Msun}{\mathrm{M}_{\odot}}
\newcommand{\Rsun}{\mathrm{R}_{\odot}}
\newcommand{\Osun}{\Omega_{\odot}}

\newcommand{\Mstar}{M\sub{\star}}

\newcommand{\Rstar}{R\sub{\star}}
\newcommand{\Mstarrate}{\dot{M}\sub{\star}}
\newcommand{\Msunrate}{\dot{M}_{\odot}}

\newcommand{\apos}{a\sub{p}}

\newcommand{\npp}{n\sub{p}}

\newcommand{\Kpp}{k\sub{2,p}}

\newcommand{\Qp}{Q\sub{p}}
\newcommand{\Kss}{k\sub{2,\star}}
\newcommand{\Qs}{Q_\star}

\newcommand{\Ip}{I\sub{p}}
\newcommand{\Is}{I\sub{\star}}
\newcommand{\Lorb}{L\sub{orb}}

\newcommand{\Der}{\mathrm{d}}

\newcommand{\porb}{P\sub{orb}}
\newcommand{\prot}{P\sub{rot\_ini, \star}}
\newcommand{\pfrot}{P\sub{rot\_final, \star}}

\newcommand{\epp}{\varepsilon\sub{p}}
\newcommand{\eps}{\varepsilon\sub{\star}}
\newcommand{\gyrp}{\zeta\sub{p}}
\newcommand{\gyrs}{\zeta\sub{\star}}
\newcommand{\omp}{\Dot{\Omega}\sub{wind, p}}
\newcommand{\oms}{\Dot{\Omega}\sub{wind, \star}}
\newcommand{\roche}{a\sub{Roche}}

\newcommand{\tdecay}{\tau\sub{decay}}
\newcommand{\tcircular}{\tau\sub{\bigcirc}}

\newcommand{\alphap}{\alpha\sub{p}}

\newcommand{\gammap}{\gamma\sub{p}}

\newcommand{\alphas}{\alpha\sub{\star}}

\newcommand{\gammas}{\gamma\sub{\star}}

\usepackage{graphicx}	
\usepackage{amsmath}	
\usepackage{amssymb}	

\title[TESS gas giants around M dwarfs]{Tidally-induced migration of TESS gas giants orbiting M dwarfs}

\author[J. A. Alvarado-Montes]{\parbox{\textwidth}{Jaime A. Alvarado-Montes$^{1, 2}$\thanks{E-mail: jaime-andres.alvarado-montes@hdr.mq.edu.au}}\vspace{0.4cm}\\
$^{1}$School of Mathematical and Physical Sciences, Macquarie University, Sydney, NSW 2109, Australia.\\
$^{2}$Research Centre for Astronomy, Astrophysics and Astrophotonics, Macquarie University, Sydney, NSW 2109, Australia.
}%

\date{Accepted 2022 September 16. Received 2022 September 15; in original form 2022 May 27}

\pubyear{2022}

\begin{document}
\label{firstpage}
\pagerange{\pageref{firstpage}--\pageref{lastpage}}
\maketitle

\begin{abstract}
According to core-accretion formation models, the conditions under which gas giants will form around M dwarfs are very restrictive. Also, the correlation of the occurrence of these planets with the metallicity of host stars is still unknown due to the intrinsic faintness of M dwarfs in the optical and some intricacies in their spectra. Interestingly, NASA's TESS mission has started to create a growing sample of these systems, with eleven observed planets located in close-in orbits: contrary to what is expected for low stellar masses. Tidal interactions with the host star will play a key role in determining the fate of these planets, so by using the measured physical and orbital characteristics of these M-dwarf systems we numerically analyse the exchange of rotational and orbital angular momentum, while constraining the energy dissipation in each system to calculate whether host stars are spun up or spun down, depending on the relationship between the gain and loss of angular momentum by the stellar rotation. We also study the coupled orbital and physical evolution of their gas giant companion and calculate orbital circularization time-scales, as well as the time needed to undergo orbital decay from their current orbital position to the Roche limit. The thorough study of tidal processes occurring over short and long time-scales in star-planet systems like those studied here, can help constrain tidal dissipation rates inside the star and planet, complement tidal theories, and improve estimations of unconstrained properties of exoplanetary systems.
\end{abstract}




\section{Introduction}
\label{sec:intro}

\begin{figure}
    \centering
    \includegraphics[scale=0.425]{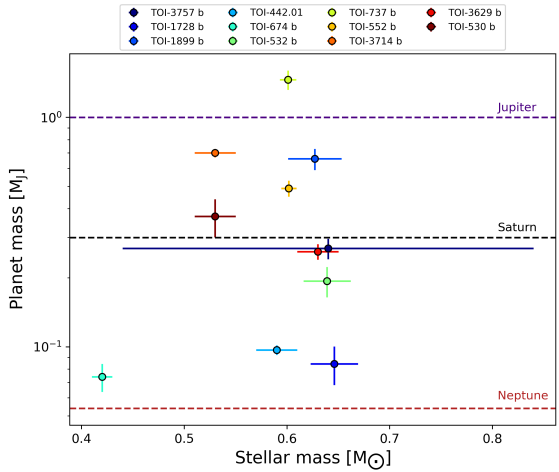}
    \caption{Stellar and planetary mass distribution of the GGM-D sample.}
    \label{fig:distri}
\end{figure}

In planet formation models based on the core-accretion framework, the formation of gas giants around M dwarfs is hampered due to the lack of solid material in the discs and the short disc lifespans owing to high UV radiation. These systems are unique: their occurrence and our ability to detect them will help us better understand how massive planets form around low-mass stars. There is a small but growing number of these planets ($\sim11$, Fig. \ref{fig:distri}; hereafter the GGM-D sample) confirmed by NASA's TESS mission \citep{Ricker2015}: TOI-1728 b \citep{Kanodia2020}, TOI-1899 b \citep{Canas2020}, TOI-442 b \citep{Dreizler2020}, TOI-674 b \citep{Murgas2021}, TOI-532 b \citep{Kanodia2021}, TOI-737 b, TOI-552 b \citep{Jordan2022}, TOI-3629 b, TOI-3714 b \citep{Canas2022}, TOI-3757 b \citep{Kanodia2022}, and TOI-530 b \citep{Gan2022}; all with close-in orbits, thus making them perfect bodies to study tidal evolution around M-dwarf hosts.

The proximity of the aforementioned planets from their host star implies a clear shift from their original orbital position (i.e. inwards migration), which might have been partially driven by tidal interactions that will now play a primary role in determining their subsequent evolution and fate. Given the short orbital period of all these planets along with their significant size and mass, the exchange of orbital angular momentum can significantly affect the rotation of the star. Interestingly, by studying in detail the mutual interactions of these systems, we may constrain the physical and orbital parameters driving tidally-induced evolution around M dwarfs.

Tidal interactions are key to understanding the orbital and physical long-term changes that close-in planets undergo \citep{Gu2004}. Such interactions have been extensively studied in the context of our Solar System \citep{Goldreich1966,Goldreich1977} and binary systems \citep{Hut1980}, and it is relatively recently that they started being extended to study the fate of compact exoplanetary systems. When giant planets orbit too close from their host stars, orbital parameters such as semi-major axis and eccentricity can be significantly modified. Depending on the stellar rotational rate, many of these planets will spiral in towards their host star and cross the Roche limit where the transfer of mass at the L1 point starts taking place \citep*{Gu2003}. At the same time, since such evolution of the planetary orbit modifies the total angular momentum of the system, the stellar and planetary rotation must change too.

The interest in how the rotation of exoplanet hosts is affected by the presence of planetary companions is growing \citep{Alves2010,Lanza2010,Leconte2010,Brown2011}. When a planet exchanges angular momentum with the star, two outcomes can be possible for the star: 1) the angular momentum gain from the planetary orbit spins up the stellar rotation, or 2) the angular momentum loss due to magnetic braking is much bigger and thus the in-falling planet simply slows down the rate at which angular momentum is lost, but it does not balance the outgoing angular momentum and the stellar rotation period reaches very low values. As we know, the latter can happen because stars progressively lose material via stellar winds \citep{Weber1967} and the rotation rate decreases asymptotically with the inverse square root of stellar age \citep{Skumanich1972}. The rotation of the two bodies under the effect of tidal interactions can provide us with invaluable information about the physical characteristics of the system; here, we study the extrasolar systems in the GGM-D sample by computing their dynamical evolution and fate.

All of the planets in the GGM-D sample, except for TOI-530 b and TOI-532 b, have a non-negligible eccentricity. The highly significant star-planet tidal effects resulting from their short orbital period will efficiently damp their eccentricity and circularise their orbit \citep*{Jackson2008a}. Also, depending upon the model adopted to study tidal interactions (see e.g. \citealt*{Alvarado2017,Alvarado2019}), eccentricity damping and orbit circularization might have important consequences for planets in eccentric orbits which may undergo orbital decay twice as fast as planets in circular orbits.

The orbital architecture of the planets in the GGM-D sample can be used to study the strength of tides in these systems and improve tidal theories thereof. Also, as this is closely related to the properties of the planet-hosting star, it could also provide important details about the stellar inner structure. Inspired by the growing number of these discoveries, we thoroughly study the tidal evolution of each planet in the aforementioned sample and, for different input parameters, constrain their tidal circularization and orbital decay time-scales.

To model properly the migration of these planets, we include the evolution of unconstrained stellar/planetary tidal-related properties. Particularly, we study the deformation of interacting bodies via: 1) the tidal Love number of second order, $k_2$,  
determined by the distortion efficiency of bodies and used to measure their response to tidal deformation; and 2) the dimensionless tidal quality function, $Q$, which stands for the efficiency to dissipate energy under a forced oscillation per tidal cycle \citep{Hansen2010}. The combination of these two parameters (i.e. $k_2/Q$), describes the dissipated energy in each system \citep{Ogilvie2013}: for M dwarfs with a close-in companion, the dissipated energy will be affected by the stellar spin and angular momentum exchanged from the planetary orbit. To study this, we adopt a two-layer interior structure model for the star \citep{Mathis2015b} and the planet \citep*{Guenel2014}; and assess the evolution of these short-period planets under a given set of orbital and physical parameters.

The layout of this paper is as follows. In Section \ref{sec:eqs}, we present our tidal model. In Section \ref{sec:analysis}, we numerically analyse the tidal evolution of each planet in the GGM-D sample and discuss the different outcomes pertaining to the parameters studied in this work. We conclude in Section \ref{sec:concl} and shortly discuss the implications of the tidal evolution of gas giants for the properties of extrasolar planetary systems.

\section{Tidal Evolution Equations}
\label{sec:eqs}

We present here the equations used to describe the tidal evolution of a single star-planet system. When interacting, both the star and the planet create mutual bulges that produce a torque conditioned by the stellar rotation rate that might shrink the planetary orbit, and drive the evolution of eccentricities and spin-orbit angles. Also, depending on the system's physical parameters, a spin-up (down) of the stellar rotational rate might be produced (see e.g. \citealt{Brown2011}). We assume here that the stellar and planetary rotation axes are aligned with the orbital plane (so-called spin-orbit alignment)\footnote{Otherwise, we need to include the evolution of the stellar and planetary obliquity (see e.g. \citealt{Barker2009}).}.

The equations presented in Section \ref{sec:orbitalevo} are the most widely adopted model due to their generality for any eccentricity. These equations result from the equilibrium tide theory introduced by \citet{Darwin1879} which uses a constant tidal lag time ($\tau$; a time proportional to the dissipation rate which is responsible for the delay of the star's tidal bulge with respect to the planet's position vector). In addition, rotating fluid bodies have complex internal mechanisms by which they dissipate energy  \citep{Zahn2008}: as a consequence, within the aforementioned model $k_2/Q$ is a complicated function of the tidal frequency $\omega$ (see e.g. \citealt{Papaloizou1997,Ogilvie2004,Ogilvie2007}). This further hinders the analysis of the secular evolution of planetary companions for different  orbital eccentricities and, in general, for arbitrary obliquities of the star and the planet. 

Given the complexity explained above, to determine the efficiency of tidal dissipation we assume that $Q$ is inversely proportional to $\omega$ so that $Q=\frac{1}{\omega\tau}$ (in the same spirit of \citealt{Barker2009}). In this regard, for the resulting secular evolution equations we adopt a constant-$Q$ model with $Q=\frac{1}{\npp\tau}$ (i.e. $\omega=\npp$, where $\npp$ is the mean motion of the planet). This implies that the lag time $\tau$ is proportional to the planet orbital period (cf. \citealt{Mardling2002}), frequency-independent, and the same for all the components of the tide.

\subsection{Orbital evolution, mass loss, and magnetic braking}
\label{sec:orbitalevo}

Following the prescription of tides coupled to stellar/planetary interior structures described in \citet{Alvarado2021}, where the assumption introduced above was not explicitly mentioned (i.e. $\tau=\frac{1}{\npp Q}$), the time evolution of the planet mean motion ($\npp$), eccentricity ($e$), spin ($\Op$), and stellar rotation ($\Os$) will be given as a set of coupled differential equations \citep{Alexander1973,Hut1981},

\beq{eq:dnpdt}
\frac{\dot{n}\sub{p}}{\npp} = -3\left[\frac{\dot{e}}{e}\frac{e^2}{1-e^2}-\frac{\Is\dot{\Omega}_\star}{\Lorb}-\frac{\Ip\dot{\Omega}\sub{p}}{\Lorb}+\frac{\Is\oms}{\Lorb}\right],
\eeq

\beq{eq:dedt}
\begin{split}
\frac{\dot{e}}{e}=\frac{27\npp}{\apos^5}&\Bigg\{\left<\frac{\Kpp}{\Qp}\right>_\mathrm{c+e}\frac{\Mstar(t)}{\Mp(t)}\Rp^5\left[\frac{11}{18}\frac{\Op}{\npp}e_2(e)-e_1(e)\right]\\&+\left<\frac{\Kss}{\Qs}\right>_\mathrm{e}\frac{\Mp(t)}{\Mstar(t)}\Rstar^5\left[\frac{11}{18}\frac{\Os}{\npp}e_2(e)-e_1(e)\right]\Bigg\},
\end{split}
\eeq

\beq{eq:dopdt}
\frac{\Der\Op}{\Der t}= \frac{3\npp^4\Rp(t)^3}{\epp\gyrp \gcons\Mp(t)}\left<\frac{\Kpp}{\Qp}\right>_\mathrm{c+e}\left[e_3(e)-e_4(e)\left(\frac{\Op}{\npp}\right)\right] + \omp,
\eeq\vspace{-0.21cm}

\beq{eq:dosdt}
\frac{\Der\Os}{\Der t}= \frac{3\npp^4\Mp(t)^2}{\eps\gyrs \gcons}\left(\frac{\Rstar}{\Mstar}\right)^3\left<\frac{\Kss}{\Qs}\right>_\mathrm{e}\left[e_3(e)-e_4(e)\left(\frac{\Os}{\npp}\right)\right] + \oms,
\eeq
with $e_1$, $e_2$, $e_3$, and $e_4$ being extension functions in $e$:

\beq{eq:ecc1}
e_1(e) = \left(1 + \frac{15}{4}e^2 + \frac{15}{8}e^4 + \frac{5}{64}e^6\right) \bigg/ (1-e^2)^{13/2},
\eeq

\beq{eq:ecc2}
e_2(e) = \left(1 + \frac{3}{2}e^2 + \frac{1}{8}e^4\right) \bigg/ (1-e^2)^{5},
\eeq

\beq{eq:ecc3}
e_3(e) = \left(1 + \frac{15}{2}e^2 + \frac{45}{8}e^4 + \frac{5}{16}e^6\right) \bigg/ (1-e^2)^6,
\eeq

\beq{eq:ecc4}
e_4(e) = \left(1 + 3e^2 + \frac{3}{8}e^4\right) \bigg/ (1-e^2)^{9/2}.
\eeq

$\Lorb$, the orbital angular momentum, is equal to
\beq{eq:lorb}
\Lorb = \Mp\Mstar\sqrt{\frac{\gcons\apos(1-e^2)}{\Mstar + \Mp}},
\eeq
where the planet semi-major axis is $\apos = [G(\Mstar + \Mp)/\npp^2]^{1/3}$. The stellar and planetary angular moment of inertia are given by
\beq{eq:moment_s}
\Is = \eps\gyrs\Mstar\Rstar^2
\eeq
and
\beq{eq:moment_p}
\Ip = \epp\gyrp\Mp\Rp^2,
\eeq
respectively. In equations (\ref{eq:moment_s}) and (\ref{eq:moment_p}), the moment of inertia coefficient $\gyrs$ ($\gyrp$) and the mass fraction $\eps$ ($\epp$) are set following \citet{Gu2003} and \citet{Dobs2004}.

The planetary mass ($\Mp$) will decrease for photo-evaporation and stellar wind drag, as does the stellar mass ($\Mstar$) due to stellar wind ($\oms$),
\beq{eq:Mloss-rate}
\Mprate = -\frac{\pi \Rp^{3} F_\mathrm{XUV}}{\gcons K\Mp(t)}- \left(\frac{\Rp(t)}{\apos}\right)^2 \frac{\Mstarrate \alpha}{2},
\eeq

\beq{eq:stellarmassrate}
\Mstarrate =\left(\frac{\Mstar}{\Msun}\right)^{a}
\left(\frac{\Os}{\Osun}\right)^{b} \left(\frac{\Rstar}{\Rsun}\right)^{2}\Msunrate.
\eeq

It is worth mentioning that both the planet mean motion evolution (equation \ref{eq:dnpdt}) and the tidal spin evolution of the star (equation \ref{eq:dosdt}) are affected by wind torques, explaining why $\oms$ appears on both equations (see \citealt{Brown2011}). All quantities in equations (\ref{eq:Mloss-rate}) and (\ref{eq:stellarmassrate}) follow \citet{Alvarado2021} and references therein: $K$ accounts for the planet radius losses at the Roche lobe, $F_\mathrm{XUV}$ is the X-rays and eXtreme UV stellar flux, $\gcons$ is the gravitational constant, $\alpha=0.3$ is an entrainment efficiency factor; $a=-3.36$, $b=1.33$, $\Osun=2.67\times10^{-6}$ rad s$^{-1}$, and $\Msunrate=1.4\times10^{-14}\Msun$ yr$^{-1}$. 

For conservative systems \citep*{Dobs2004}, planetary winds can be neglected and $\omp\simeq0$. However, due to magnetic braking (i.e. loss of stellar material) the host stars lose angular momentum \citep{Weber1967,Skumanich1972}, and $\oms$ will be calculated as
\beq{eq:magbrak}
\oms=-\kappa \Os \mathrm{Min}(\Os, \Obar)^2.
\eeq

In equation (\ref{eq:magbrak}), $\Obar$ depends on the stellar spectral type and denotes the `saturation' rate, which represents the upper limit where the magnetic dynamo no longer depends on the stellar spin (so-called the `saturated' regime); $\kappa$ determines how the magnetic braking rate scales physically. To compute $\kappa$, we take the change in stellar rotation due to the stellar wind as $\Dot{\Omega}\sub{\star}=-\kappa\Os^3$ \citep{Weber1967}. Integrating as $\int_{\Omega\sub{\star, 1}}^{\Omega\sub{\star, 2}}\Os^{-3}\Der\Os = -\kappa \int_{t_1}^{t_2}\Der t$, and if $\Omega\sub{\star, 2}\ll\Omega\sub{\star, 1}$ and $t_2\gg t_1$, we assume that $\Omega\sub{\star, 2}\approx\Osini\approx\Os$ and $t_2\approx \tage$, where $\Osini$ is the stellar rotational rate prior to orbital decay and $\tage$ is the age of the star. Following this, $\kappa$ is computed as follows
\beq{eq:kappa}
\kappa = \frac{\Os^{-2}}{2\tage},
\eeq
showing that the torque produced by magnetic braking inversely depends on the age and spin rate of the star (see e.g. \citealt{Reville2015a, Reville2016a, Finley2017}).

\vspace{-5pt}
\subsection{Calculating $\lowercase{k}_2/Q$}
\label{sec:k2q}
The overall tidal evolution of a close-in planetary system (i.e. equations \ref{eq:dnpdt}, \ref{eq:dedt}, \ref{eq:dopdt}, and \ref{eq:dosdt}) depends on how the function of tidal energy dissipation ($k_2/Q$) is calculated over time. A contributor to tidal dissipation in stars with convective envelopes is the dissipation of equilibrium tides in convection zones. Equilibrium tides travelling through convective regions will encounter turbulent motions that produce a net effect as that of an effective viscosity. M dwarfs can have slow convective velocities which can reduce the convective frequency $\omega_c$, meaning that for the same tidal period the ratio $\omega/\omega_c$ is larger and the turbulent viscosity is smaller (even without reducing the frequency). Therefore, in M dwarfs, despite the great depth of their convection zones (and high density regions), the convective damping of equilibrium tides is weaker than in solar-mass stars.

In a given stellar model, if we want to study the damping of tidal flows we need to set the turbulent effective viscosity, for example, assuming an isotropic viscosity (see e.g. \citealt*{Penev2009}), which can be done if the dependence of the tidal frequency on viscosity is known. Also, the effective turbulent viscosity in M dwarfs that describes the interaction between convection and equilibrium tidal flows, strongly depends on the frequency. Such a dependence is not clear and various approaches have recently been considered \citep*{Duguid2020a,Duguid2020b}, although no agreement is still defined about how equilibrium tides are dissipated due to the interaction between tidal flows and turbulent convection.

The viscosity-dependence of the equilibrium tides damping in convective regions is significant and depends on the tidal period. In this regard, adopting different prescriptions for the turbulent viscosity will lead to different values of the tidal function $k_2/Q$ due to equilibrium tides, thus changing the effect of equilibrium tides on the tidal evolution of a given M-dwarf system. For short-period planets around low-mass and solar-mass stars, the effective viscosity in the envelope is significantly reduced \citep{Barker2020}. For small values of the tidal frequency, $k_2/Q$ due to equilibrium tides is very small (i.e. $\sim 10^{-12} - 10^{-11}$) when compared to that of inertial waves and/or internal gravity waves. In fact, for high orbital frequencies (small periods) there is a big uncertainty about the correct model to describe the effective turbulent viscosity in convection zones, so for different models energy dissipation can differ up to 3 orders of magnitude, while also values already calculated for $k_2/Q$ could be overpredicted from 1 to 3 orders of magnitude \citep{Barker2020}

Due to the reasons above, for the M-dwarf systems of this work we decide to neglect equilibrium tides when calculating the dissipation of energy inside the star and the planet. The dissipation of energy inside the star will be produced by dynamical tides, namely both the excitation of inertial waves (IWs) in the convective envelope and internal gravity waves (IGWs) in radiative regions. For the planet, dissipation in the convective envelope is also produced by IWs whereas a viscoelastic model is used for the dissipation in the inner solid core.

\subsubsection{Dissipation due to Inertial Waves in Convective Envelopes}
\label{sec:k2q_IW}

Interacting bodies can change their equilibrium shape due to tidal interactions. If the amplitude of those tidal deformations are small, such changes can be assumed proportional to the distorting forces \citep{Love1927}. How the planet responds to tidal stresses can be measured by using the so-called Love numbers, $k_l^m(\omega)$, which are complex coefficients that depend on the tidal frequency $\omega$ (see e.g. \citealt{Efroimsky2012}). If we assume negligible obliquities for the star and the planet, the imaginary part of $k_2^2(\omega)$ averaged in $\omega\,\epsilon\,[-2\Omega, 2\Omega]$ will provide us information about the tidal dissipation rate due to IWs,

\beq{eq:Imk2}
\frac{k_2}{Q} =\int_{-\infty}^{+\infty}{\rm Im}[k_2(\omega)]\frac{\Der\omega}{\omega}=\int_{-\infty}^{+\infty}\frac{|k_2^2(\omega)|}{Q_2^2(\omega)}\frac{\Der\omega}{\omega}.
\eeq

We adopt here a simplified interior structure model of two layers for both the star and the planet, and we are using a frequency-averaged model of tidal dissipation due the excitation of IWs in convective envelopes of rapidly rotating stars and planets. This mechanism will only apply when $\omega\leq2\Os$, which for small eccentricities will be true only if $\porb\geq\frac{\prot}{2}$. Also, by opting for such a model we are filtering out the frequency dependence of tidal dissipation in spherical shells \citep{Ogilvie2007}: this might lead to under- or over-estimation of tidal dissipation rates.

It is worth noting that a constant time-lag and $k_2/Q$ models have different functional forms of the planet semi-major axis. In reality, neither of them is strictly correct, and for small eccentricities as those studied throughout this paper, assuming a frequency-averaged dissipation would produce quantitative differences rather than qualitative. However, given the complexity to compute tidal dissipation for realistic stars and giant planets (due to their dependence on $\omega$), we found that these assumptions are the best way to move forward and do a population study by analysing the general effects of tidal friction (due to IWs) on planetary orbital evolution.

For the stars (subscript $\star$), a fluid-fluid boundary is used between the two layers as shown in \citet{Mathis2015b}; for the giant planets (subscript p), we use a solid-fluid boundary as in \citet{Guenel2014},
\beq{eq:k2QFormulas}
\begin{split}
    \left<\frac{\Kss}{\Qs}\right>_\mathrm{e} =& \frac{100\pi}{63}\epsilon_\mathrm{\star}^{2}\frac{\alphas^{5}}{1-\alphas^{5}}(1-\gammas^2)(1-\alphas^2)\times\\&\left(1 + 2\alphas + 3\alphas^2 + \frac{3}{2}\alphas^3\right)^2\left[1+\left(\frac{1-\gammas}{\gammas}\right)\alphas^3\right]\times\\&\left[1 + \frac{3}{2}\gammas+\frac{5}{2\gammas}\left(1 + \frac{1}{2}\gammas-\frac{3}{2}\gammas^2\right)\alphas^3-\frac{9}{4}(1-\gammas)\alphas^5\right]^{^{-2}}
\end{split}
\eeq
\beq{eq:k2QFormulap}
\begin{split}
    \left<\frac{\Kpp}{\Qp}\right>_\mathrm{e+c} =& \frac{100\pi}{63}\epsilon_\mathrm{p}^{2}\frac{\alphap^{5}}{1-\alphap^{5}}\left[1+
	\frac{1-\gammap}{\gammap}\alphap^{3}\right]\times\\&\left[1+
		\frac{5}{2}\frac{1-\gammap}{\gammap}\alphap^{3}\right]^{^{-2}} + \frac{\upi\mathcal{R}(3 + \mathcal{A})^{2}\mathcal{BC}}{\mathcal{D}(6\mathcal{D} + 4\mathcal{ABCR})},
\end{split}
\eeq 
where $\epsilon^{2}\equiv (\Omega/\Ocri)^{2}$ and $\Ocri\equiv(\gcons M/R^{3})^{1/2}$. The first and second term in equation (\ref{eq:k2QFormulap}) represent the dissipation via inertial waves in the envelope and viscoelastic dissipation in the planet inner solid core, respectively. The core's rigidity, $\mathcal{R}$, is measured in Pascals [Pa], while $\mathcal{A}$, $\mathcal{B}$, $\mathcal{C}$, and $\mathcal{D}$ are auxiliary definitions given by
\beq{eq:corepars}
\begin{split}
&\mathcal{A} = 1 + \frac{5}{2}\gamma^{-1}\alpha^3(1-\gamma), \quad\mathcal{B} = \alpha^{-5}(1 - \gamma)^{-2},\\&
\mathcal{C} = \frac{38\upi}{3}\frac{(\alpha\Rp)^4}{\gcons(\beta\Mp)^2},\hspace{0.12cm}\mathcal{D} = \frac{2}{3}\mathcal{A}\mathcal{B}(1-\gamma)\left(1+\frac{3}{2}\gamma\right) - \frac{3}{2}.
\end{split}
\hspace{0.8cm}
\eeq

In equations (\ref{eq:k2QFormulas}), (\ref{eq:k2QFormulap}), and (\ref{eq:corepars}), $\alpha$ and $\beta$ are defined as the size and mass ratios in terms of the core's mass ($\Mc$) and radius ($\Rc$):
\beq{eq:k2Qparameters}
\alpha \equiv \frac{R_\mathrm{c}}{R(t)},\quad
\beta \equiv \frac{M_\mathrm{c}}{M(t)},\quad
\gamma \equiv \frac{\alpha^{3}(1-\beta)}{\beta(1-\alpha^{3})}.
\hspace{1.25cm}
\eeq

\begin{figure}
    \centering
    \includegraphics[scale=0.4]{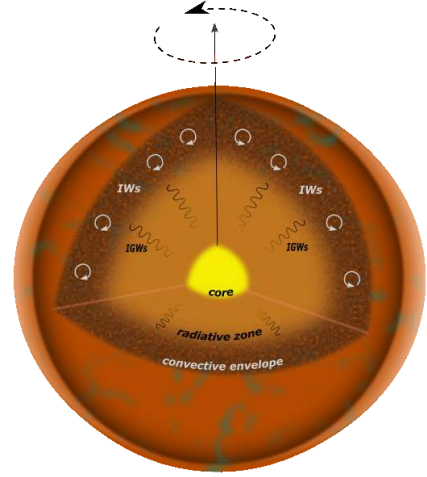}
    \caption{Schematic diagram of the interior model of the M dwarfs studied in this work. The two mechanisms studied here, namely inertial waves and internal gravity waves, are represented in white and black, respectively.}
    \label{fig:scheme}
\end{figure}

\subsubsection{Dissipation due to Internal Gravity Waves in Radiative Cores}
\label{sec:k2q_IGW}
Tidal dissipation inside the convective envelope due to excitation of IWs (subsection \ref{sec:k2q_IW}) is the dominant mechanism when $\omega\leq2\Os$ holds true, but otherwise the orbital evolution of planetary companions is not efficient to activate this mechanism. Consequently, for this mechanism to operate, stars that are fast rotators (e.g. $\prot=5$ d) would need planets in moderately short-period orbits (e.g. $\porb=2.5$ d), and conversely, IWs in slowly rotating stars would only be excited by planets in longer orbits that might even fall outside the `short-period' classification (as it will be the case for some of the planets in the GGM-D sample).

However, within the simplified two-layer model assumed in this work, stars are composed by a convective envelope but also by a radiative core: thereby IWs excited by tidal forcing are not the only relevant dissipation mechanism. In fact, we assume that the convective/radiative interface produces IGWs that travel and, in between resonances, this `travelling regime' provides us with the most efficient tidal dissipation due to IGWs. This will only apply if IGWs are fully damped in the radiative zone, which can occur due to large tidal amplitudes as well as, for example, radiative diffusion. Either mechanism will prevent IGWs from reaching the centre of the star, or to be bounced back from a different boundary.

As mentioned in \citet{Barker2020}, tidal dissipation due to IGWs in radiative zones strongly depends on the orbital/tidal period. Particularly, after complex numerical calculations coupled to stellar evolution models using \textsc{MESA}, they found a relationship between $k_2/Q$ and $\porb$ for stars from 0.5 to 1.1 $\Msun$,

\beq{eq:k2q_IGW}\left<\frac{\Kss}{\Qs}\right>_\mathrm{c,\,IGWs}^{-1}\approx \frac{2}{3}[1, 3]\times10^5\left(\frac{\porb}{0.5\mathrm{d}}\right)^{\frac{8}{3}},\eeq
where we have used $Q'=\frac{3Q}{2k_2}$, to write everything in terms of $\frac{k_2}{Q}$ which is the notation used throughout this paper.

To study the effect of IGWs on the orbital decay of the planets in the GGM-D sample, and given that 90\% of the host stars are within the mass interval for which equation (\ref{eq:k2q_IGW}) is valid, we use this equation to study tidal dissipation when IWs are not excited in the convective envelope (i.e. for slowly rotating stars). For rapidly rotating stars where $\omega\leq2\Os$, both IWs and IGWs will act simultaneously, but once this condition is not longer satisfied, for example, if the planet orbital period decreases significantly, then IGWs will be the dominant dissipation mechanism for the shortest orbital periods.

Still, we must state some specific caveats in the treatment of IGWs that will affect the analysis of tidal dissipation for the planets studied in this paper. As explained by \citet{Barker2020}, IGWs are more efficiently damped when wave breaking takes place, which happens for sufficiently large tidal amplitudes and also for large resonant tidal forcing \citep{Barker2010,Barker2011}. That being said, wave breaking will take place if waves overturn the stratification (please see \citealt{Barker2010} and \citealt{Barker2020} for a numerical evaluation of this criterion) which only occurs for a combination of stellar age, stellar mass, and planetary mass. This happens when planets exceed a mass threshold defined as the critical mass ($M_\mathrm{crit}$), which depends on the stellar structure\footnote{To analyse how $M_\mathrm{crit}$ varies with stellar mass and age, the reader can refer to fig. 9 in \citet{Barker2020}.}. According to this, the planets in the GGM-D sample would not necessarily fulfil the requirement to damp IGWs in the radiation zone and, instead, IGWs might be fully damped in a later stage of planetary orbital evolution when the critical mass for wave breaking decreases to lower values as stars get older. Also, given that most stars in the GGM-D sample are slow rotators, we are ignoring rotation in IGWs (which affects $\omega$) since their numerical treatment (see e.g. \citealt{Ogilvie2007}; \citealt{Ivanov2013}) is out of the scope of this work.

\section{Numerical Analysis}
\label{sec:analysis}

We numerically solved all the equations presented in Section \ref{sec:eqs}, using an integrator (written in \textsc{python}) that alternates between a nonstiff- and stiff-method according to the dynamical evolution of the system. All the planets in the GGM-D sample have initial positions beyond their Roche radius, $\roche$, so their self-gravity is still larger than the stellar tidal stresses \citep{Roche1849}. Their orbital evolution is computed until crossing $\roche$, which is given by
\beq{eq:rtidal}
\roche = \eta\left(\frac{\Mstar}{\Mp}\right)^{1/3}\Rp,
\eeq
where $\eta\simeq2.7$ from simulations of disrupted hot Jupiters \citep*{Guillochon2011}.

Also, in equation (\ref{eq:magbrak}), due mainly to the uncertainty of $\Obar$ (as it needs empirical estimations), for all the M-dwarf hosts in the GGM-D sample we use the nominal values for $\Obar$ provided by \citet{Cameron1994}, which cover the range of masses studied in this work. In equations (\ref{eq:dnpdt}), (\ref{eq:dedt}), (\ref{eq:dopdt}), and (\ref{eq:dosdt}) $k_2/Q$ follows the framework presented in subsection \ref{sec:k2q}, and since the stellar interiors of these stars are not well-constrained, $\alpha$ and $\beta$ (equation \ref{eq:k2Qparameters}) are set using \citet{Gallet2017} where different values for a wide range of stellar masses are provided.

Regarding the planetary interior, we must stress that assuming a piecewise-homogeneous two-layer model is an important simplification of the problem in hand (see subsection \ref{sec:k2q}). While for the planetary envelope (first term in equation \ref{eq:k2QFormulap}) we use the values adopted in \citet{Guenel2014}, we must simplify the treatment for the planetary core represented by the second term in the same equation. For the planets in the GGM-D sample, we are going to assume Jupiter-like values as a reference for the core's rigidity, using $\mathcal{R}=4.46\times10^{10}$ Pa which matches the tidal dissipation of Jupiter at the tidal frequency of Io \citep{Lainey2009,Lainey2012}. However, recent studies of Jupiter \citep{Wahl2017,Debras2019} that use data from the Juno mission \citep{Bolton2017} have shown that Jupiter's core is not completely solid and that its innermost part has an extended region very rich in heavy elements. This makes Jupiter's core diluted \citep{Muller2020}, differing from the three-layer structure that is commonly used in various models. 

The GGM-D sample presents two groups, with unknown (6 planets) and known (5 planets) initial stellar rotation periods ($\prot$). This allows us to study two different parameters, that is $\prot$ (Section \ref{sec:unknown}) and $\eps\gyrs$ (Section \ref{sec:known}), as both play a key role in the exchange of angular momentum, circularization time-scales, spiral-in of the planet, and stellar spin-up (down). To analyse the evolution of each gas giant (within each subgroup), we run five hundred simulations per planet for each parameter being studied. All other parameters are fixed following the literature of each planet, always adopting the central value because the upper/lower limits have a negligible effect on the system's overall tidal evolution.

\begin{figure*}
    \centering
    \subfigure[]{%
    \label{fig:toi1728}
    \includegraphics[scale=0.38]{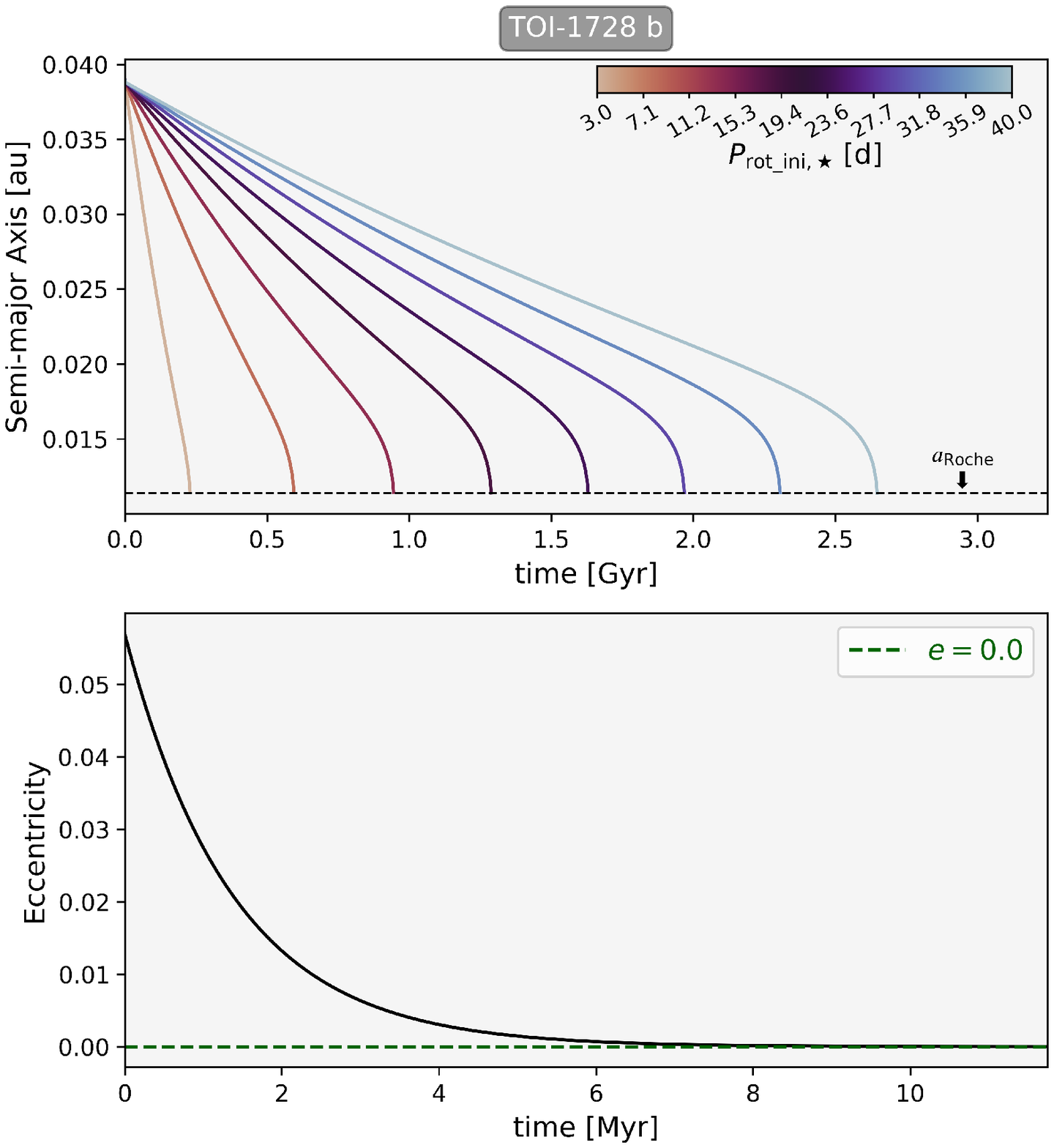}}%
    \qquad
    \subfigure[]{%
    \label{fig:toi532}
    \includegraphics[scale=0.38]{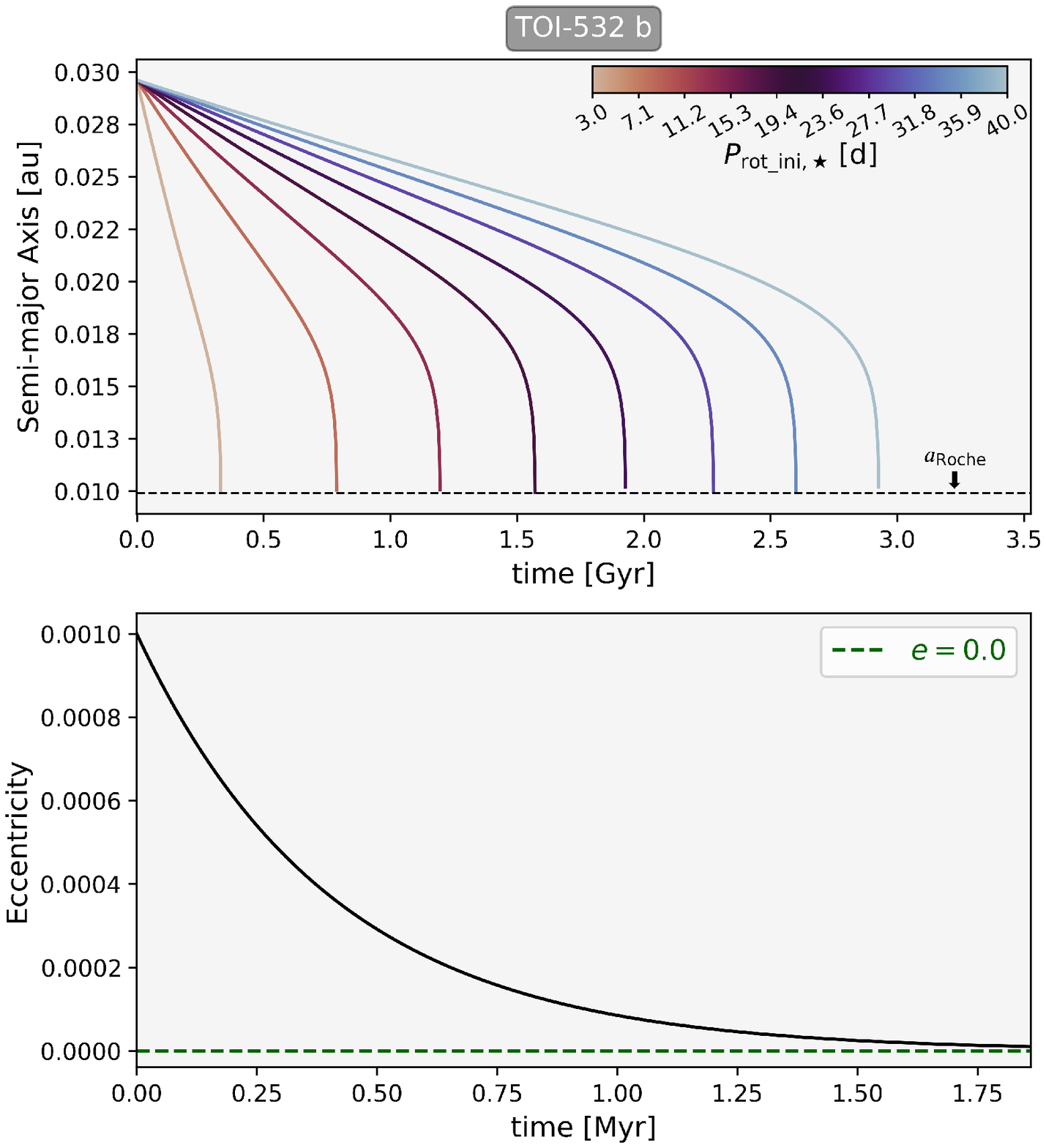}}%
    \qquad
    \subfigure[]{%
    \label{fig:toi3757}
    \includegraphics[scale=0.38]{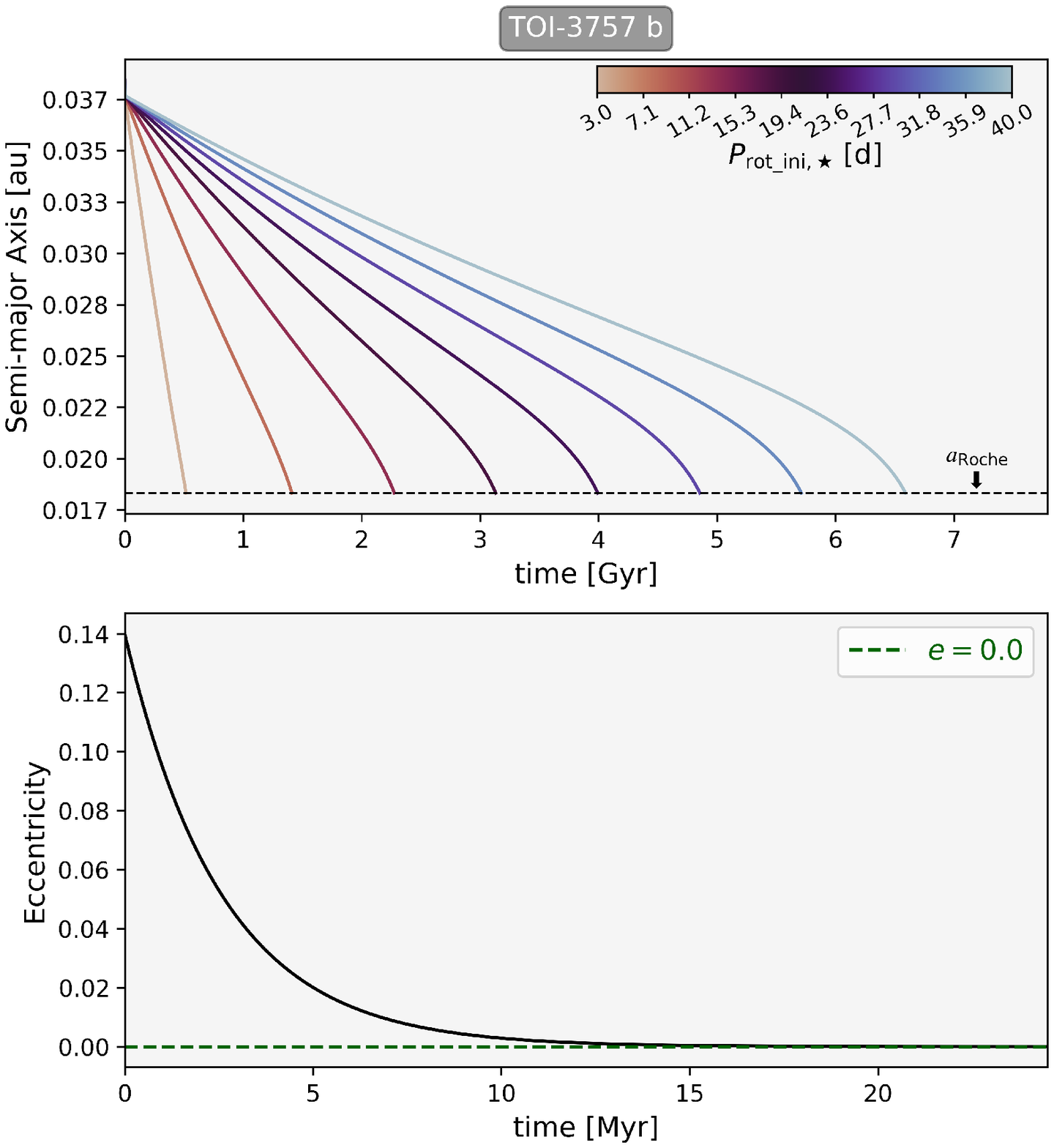}}%
    \qquad
    \subfigure[]{%
    \label{fig:toi3629}
    \includegraphics[scale=0.38]{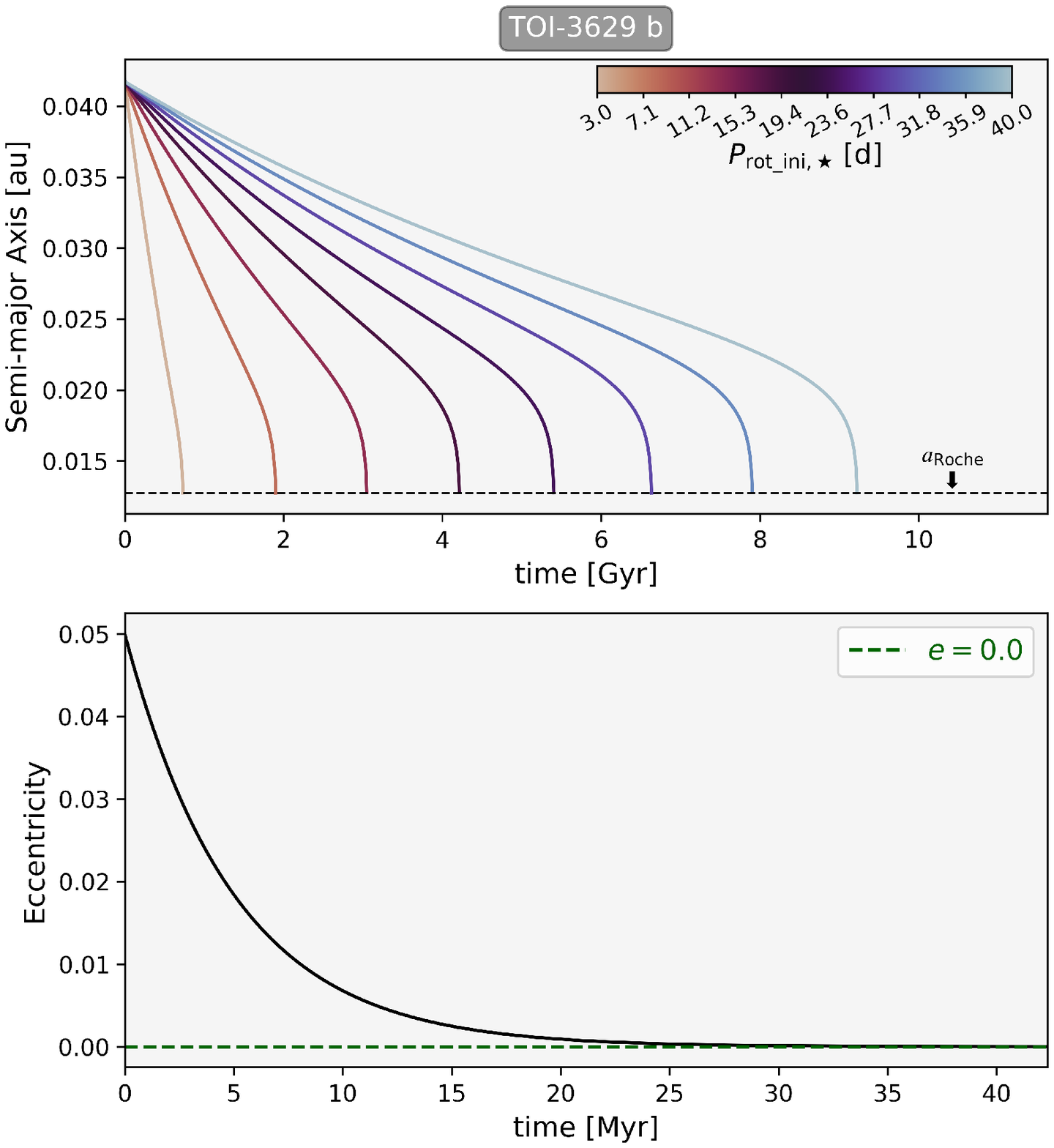}}%
    \qquad
    \subfigure[]{%
    \label{fig:toi530}
    \includegraphics[scale=0.38]{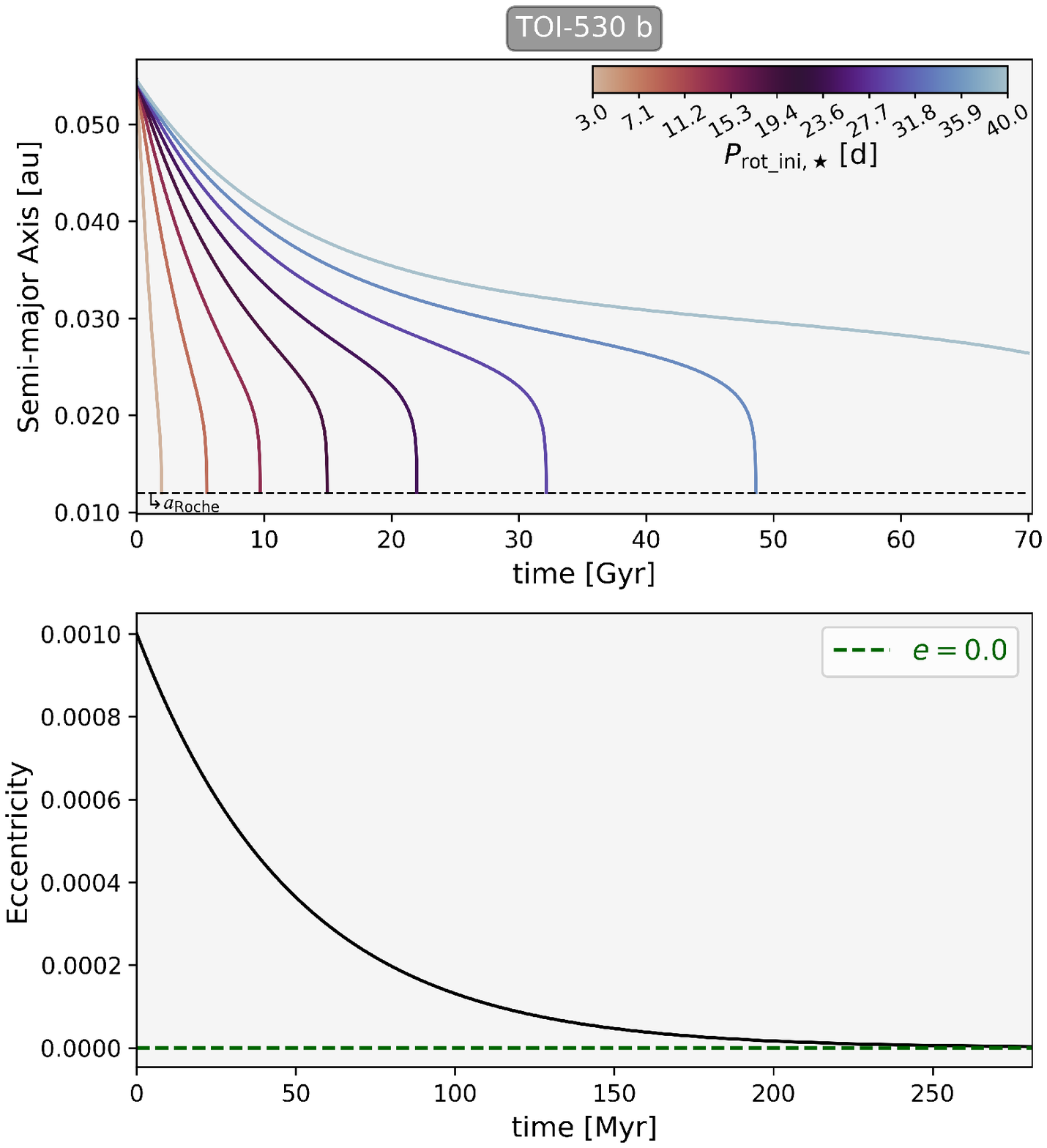}}%
    \qquad
    \subfigure[]{%
    \label{fig:toi1899}
    \includegraphics[scale=0.38]{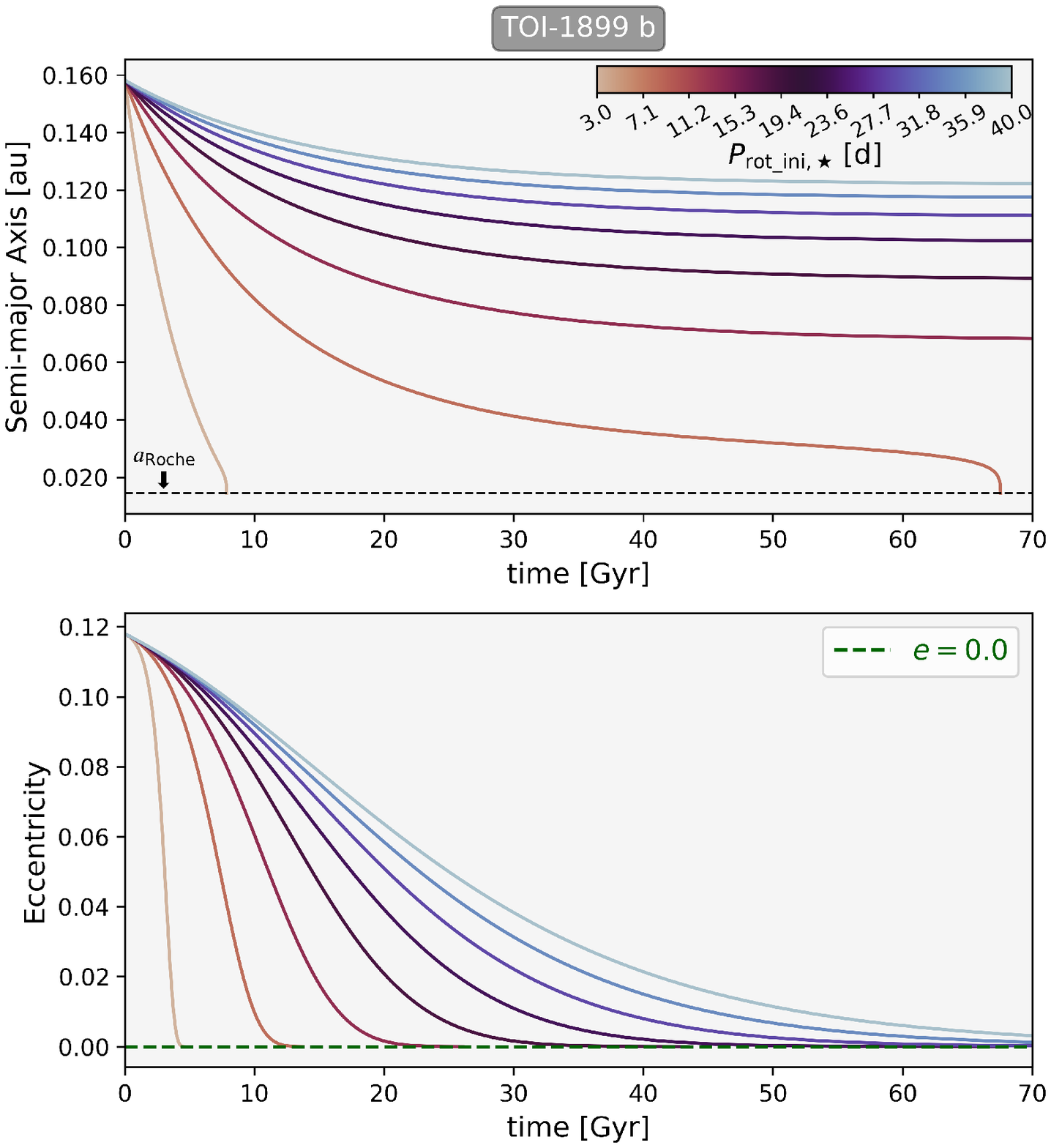}}%
    \caption{Planetary semi-major axis (upper sub-panels) and eccentricity (lower sub-panels) as a function of time, for the M-dwarf systems: TOI-1728 (a), TOI-532 (b), TOI-3757 (c), TOI-3629 (d), TOI-530 (e), and TOI-1899 (f).}
    \label{fig:unknown}
\end{figure*}

For repeatability, our numerical code uses the integrator \textsc{odeint} from \textsc{odepack} (written in \textsc{fortran}), which is publicly available. This integrator uses a method called LSODA\footnote{\url{https://github.com/scipy/scipy/blob/v0.19.0/scipy/integrate/odepack/readme}} for explicit ODEs like the ones in Section \ref{sec:eqs}, alternating between Adams and BDF methods according to the evolution of the system. The time-step was chosen based on both numerical accuracy and computational efficiency. We increased the time-step so that smoothness between consecutive times was significantly reduced (i.e. upper  time-step), and from there we started decreasing the time-step until accuracy was achieved but with the code being computationally efficient (lower time-step). For this work, we chose time-steps ranging from 1 to 10 kyr, however, some simulations were also run with finer time-steps finding no difference in numerical results but a significant increase in computational time.

\subsection{Unknown stellar rotation periods}
\label{sec:unknown}

The tidally-induced migration of these planets was studied for $3.0\leq\prot\leq40.0$ d (see Fig. \ref{fig:unknown}). The simulations of this section were run using $\eps\gyrs=0.5$. The orbital decay time-scales ($\tdecay$) and circularization times ($\tcircular$) for this subgroup are reported in Table \ref{tab:unknown}. Figs \ref{fig:toi1728}, \ref{fig:toi532}, \ref{fig:toi3757}, \ref{fig:toi3629}, and \ref{fig:toi530} correspond to TOI-1728 b, TOI-532 b, TOI-3757 b, TOI-3629 b, and TOI-530 b, respectively. In that order, $\tdecay$ has an increasing behaviour due to their combination of initial orbital period and eccentricity, which are the overarching parameters responsible for migration induced by stellar tides. 

For the systems mentioned above, the combination between $\porb$ and $\prot$ makes IWs very unlikely to be the mechanism that would drive the evolution of the planetary semi-major axis, at least for those systems with large $\prot$. For TOI-1728 b, TOI-532 b, TOI-3757 b, TOI-3629 b, and TOI-530 b the excitation of IWs in convective envelopes is the dominant mechanism contributing to tidal dissipation only when $\prot$ is smaller than 6.98, 4.65, 6.88, 7.87, and 12.76 d, respectively. That being said, equation (\ref{eq:k2QFormulas}) is only applied when the bound $\porb>\frac{\prot}{2}$ is satisfied, since IWs are only excited within such a regime (see \citealt{Barker2020}). On the other hand, as we are assuming synchronised non-circular orbits, IWs will be excited because the relevant tidal frequency and the rotation frequency of the planet have the same magnitude. Then, equation (\ref{eq:k2QFormulap}) is likely to be applied during the whole tidal evolution.

\begin{figure}
    \hspace{-0.2cm}
    \includegraphics[scale=0.69]{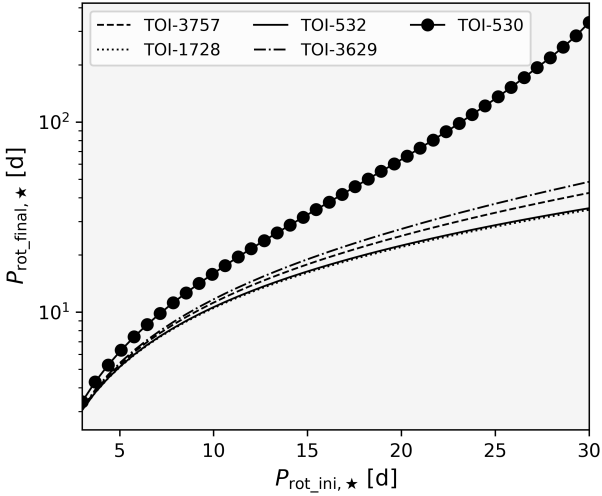}
    \caption{Final stellar rotation periods as a function of the initial stellar rotation periods, for five of the planets in subsection \ref{sec:unknown} with unknown initial rotation periods (see Table \ref{tab:unknown}).  This plot is only shown until $\prot=30$ d.}
    \label{fig:rotation_ini}
\end{figure}

When $\prot$ is larger than the aforementioned values, the dominant mechanism was the damping of IGWs in radiative zones. Given that these five systems have short orbital periods, the dissipation due to this mechanism is progressively more intense as the planetary orbit shrinks until the Roche limit. From the numerical simulations, we found that for small $\prot$ the rotation of TOI-1728, TOI-532, TOI-3757, TOI-3629, and TOI-530 did not change significantly, and the final stellar rotation period differed by approximately 5\% with respect to $\prot$. For $\prot$ longer than the bound limits mentioned above, such host stars were spun down and the magnitude of this change was directly impacted by the magnitude of $\prot$; that is, the longer $\prot$ the slower the final rotational rate of the star.

\begin{figure*}
    \centering
    \subfigure[]{%
    \label{fig:toi442}
    \includegraphics[scale=0.38]{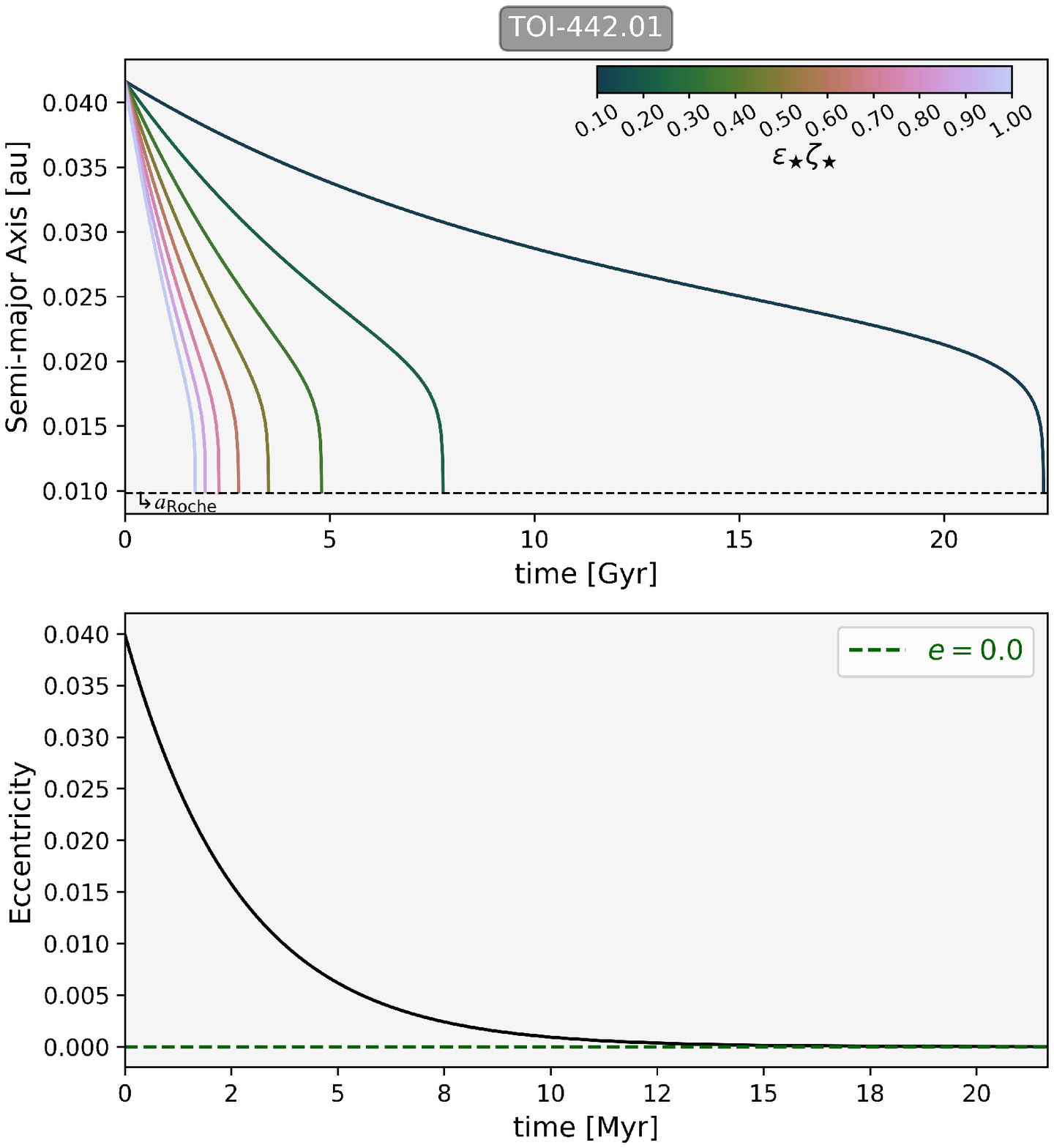}}%
    \qquad
    \subfigure[]{%
    \label{fig:toi552}
    \includegraphics[scale=0.38]{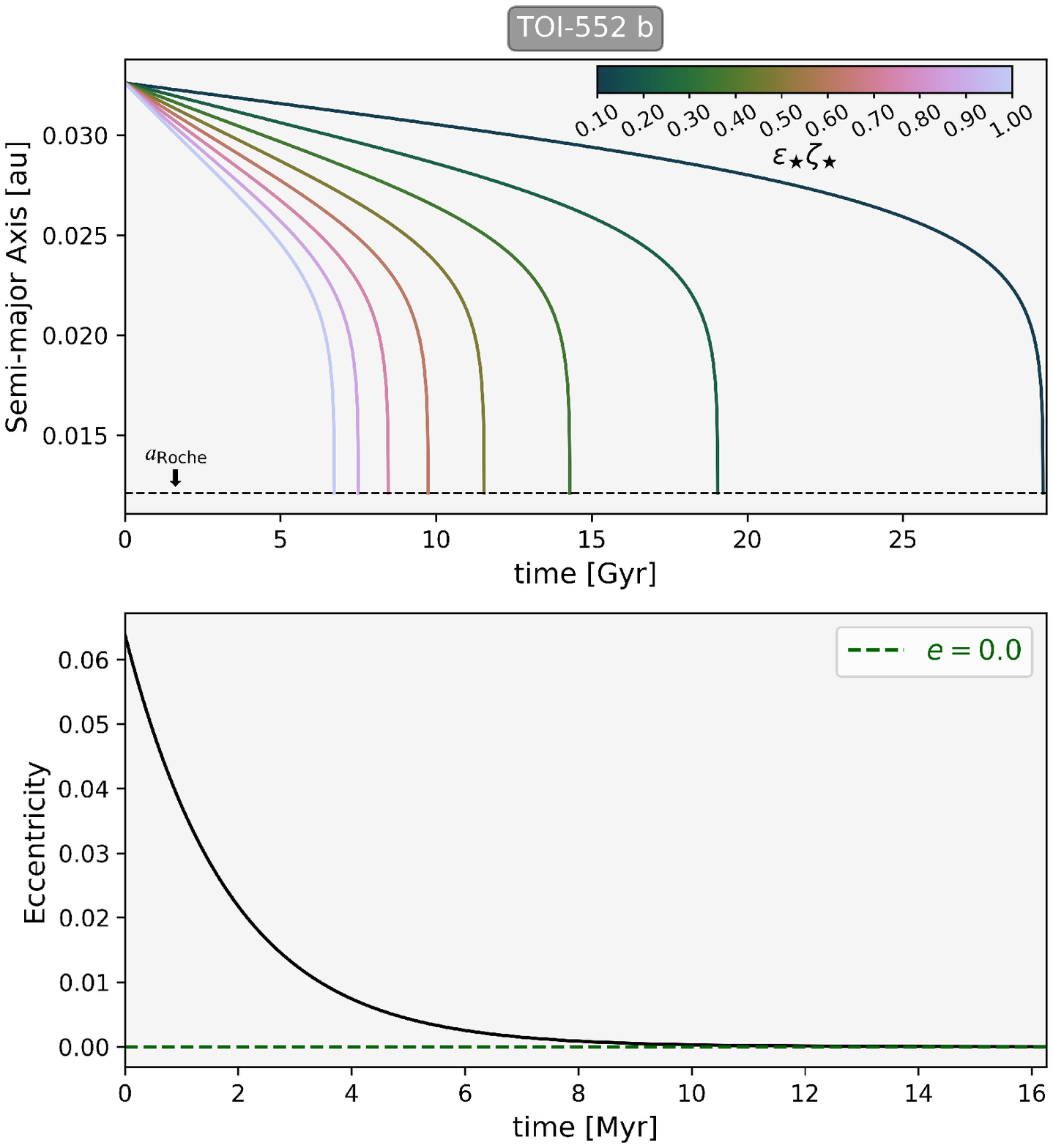}}%
    \qquad
    \subfigure[]{%
    \label{fig:toi674}
    \includegraphics[scale=0.38]{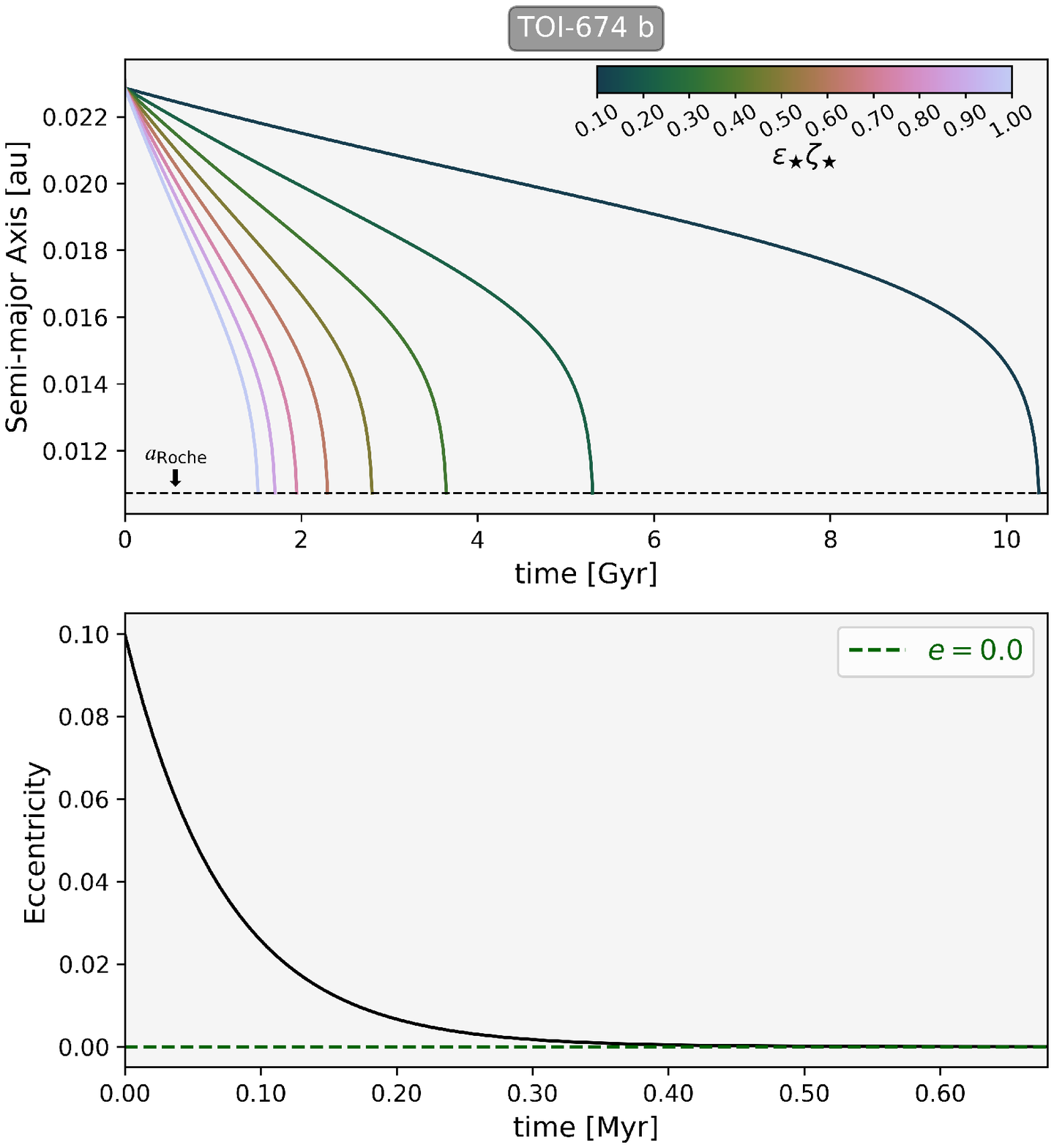}}%
    \qquad
    \subfigure[]{%
    \label{fig:toi3714}
    \includegraphics[scale=0.38]{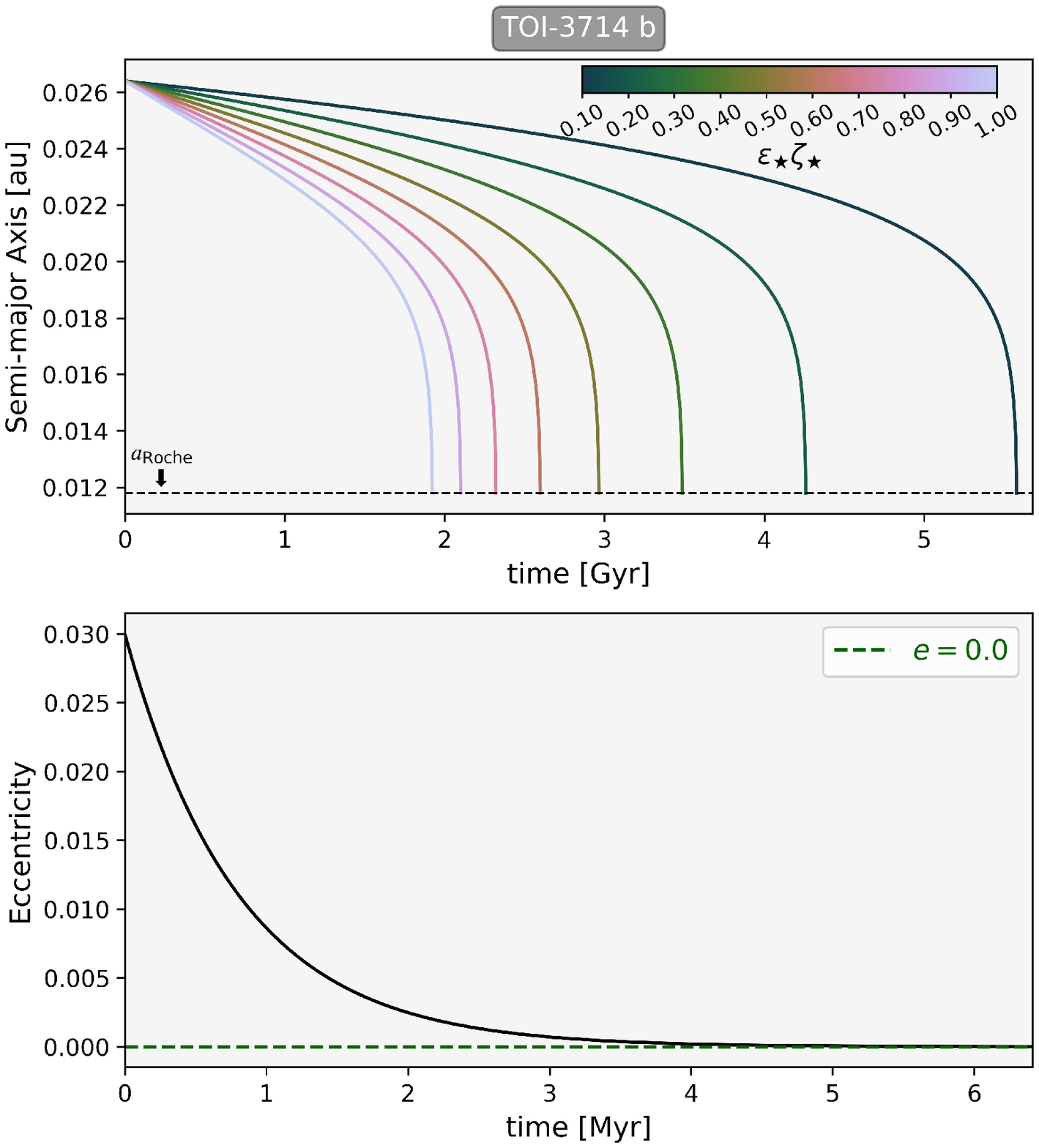}}%
    \qquad
    \subfigure[]{%
    \label{fig:toi737}
    \includegraphics[scale=0.38]{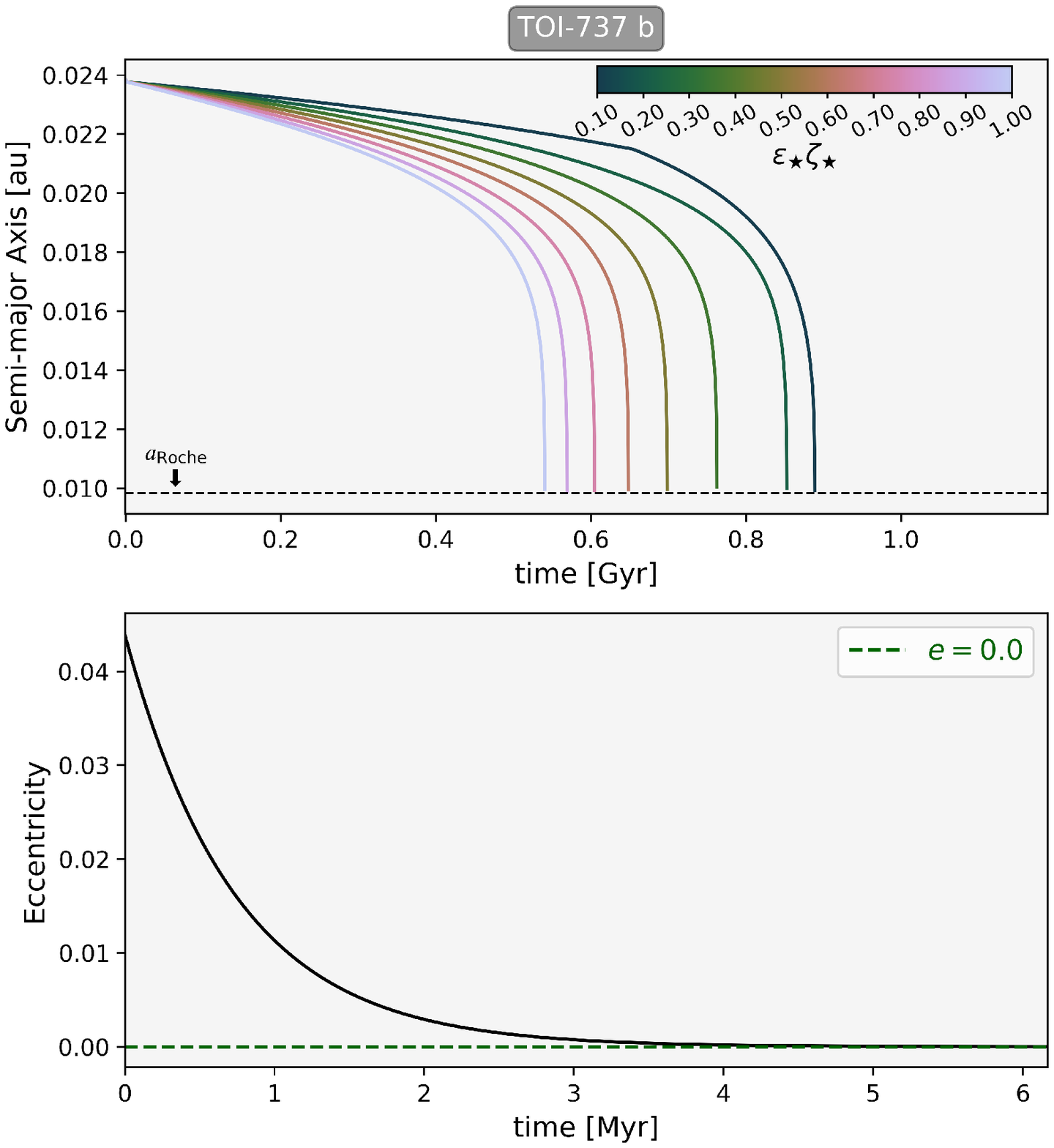}}%
    \caption{Same as Fig. \ref{fig:unknown}, but for the M-dwarf systems: TOI-442.01 (a), TOI-552 (b), TOI-674 (c), TOI-3714 (d), and TOI-737 (e).}
    \label{fig:known}
\end{figure*}

In Fig. \ref{fig:rotation_ini} we show the final stellar rotation period ($\pfrot$) in terms of  $\prot$, as computed from the simulations for TOI-1728, TOI-532, TOI-3757, TOI-3629, and TOI-530. The difference in the maximum final stellar rotation period among the planets is due to the combination of orbital period, eccentricity, and stellar/planetary mass for each system. However, the most important effect is that of the orbital period: the farther the planet is initially located, the weaker the tidal interactions with the host star and the smaller the transfer of angular momentum from the orbit to the stellar spin. Therefore, the rotation of the star decelerates due to the loss of material via stellar wind reaching lower values than a system with a close-in planet.

For TOI-1728, TOI-532, TOI-3757, TOI-3629, and TOI-530 the loss of angular momentum due to magnetic braking is evident from the low values of $\pfrot$, and there is no scenario where the stellar rotation was spun up. As expected, stellar wind was the main source of angular momentum losses and the angular momentum transferred from the planetary orbit was not enough to achieve a spin-orbit equilibrium (i.e. $\Os\simeq\npp$), so planet continued migrating inwards until crossing the Roche limit and eventually merging with their host star \citep{Dobs2004}. For these four planets, the rate at which they decay until $\roche$ is a decreasing function of $\prot$ because $\Os<\npp$ in all cases, and so planets do not survive.

Except for TOI-1899 b, the planets in this subgroup undergo orbital decay for the $\prot$ that were studied, due mainly to the damping of IGWs (since excitation of IWs does not occur in the star when stellar rotation is sufficiently low), showing a difference in $\tdecay$ of approximately one order of magnitude between the shortest and longest $\prot$. Such a difference arises due to the exchange of angular momentum: the faster the stellar rotational rate (short $\prot$) the more efficient the exchange of angular momentum. And conversely, if the star spins too slow (long $\prot$) the transfer of angular momentum to the planetary orbit is less efficient. Similarly, all of these planets (except for TOI-1899 b) circularise their orbit in time-scales ($\tcircular$) of the order of Myr, while $\tcircular$ is about three orders of magnitude larger (Gyr) for TOI-1899 b.

\setlength{\tabcolsep}{5.5pt}
\begin{table}
\centering
    \begin{tabular}{c|c|c|c|c}
         \hline\hline planet & $e$ & $\porb$ [d] & $\tdecay$ [Gyr] & $\tcircular$ [Myr] \\ \hline
         TOI-1728 b & 0.057 & 3.49151 & 0.23 -- 2.65 & 11.75 \\
         TOI-532 b & 0.001$^*$ & 2.3266508 & 0.33 -- 2.93 & 1.86\\
         TOI-3757 b & 0.14 & 3.438753 & 0.52 -- 6.59 & 24.5\\
         TOI-3629 b & 0.05 & 3.936551 & 0.73 -- 9.22 & 42.33\\
         TOI-530 b & 0.001$^*$ & 6.38758 & 1.95 -- >70 & 224.07\\
         TOI-1899 b & 0.118 & 29.02 & 7.88 -- >70 & 4.61 -- >70 [Gyr]\\
         \hline
    \multicolumn{4}{l}{\small $^*$This is a fixed value.}\vspace{-0.02cm}
    \end{tabular}
    \caption{Gas giants in subsection \ref{sec:unknown}. Here, $\tdecay$ and $\tcircular$ are from this work, whereas $e$ and $\porb$ are extracted from the references in Section \ref{sec:intro}.}
    \label{tab:unknown}
\end{table}

For TOI-1899 b, $\porb$ is greater than $\frac{\prot}{2}$ for all $\prot$ studied in this section. As a result, IWs in the convective stellar envelope were the dominant mechanism for the evolution of the planetary orbit. From this subgroup of the GGM-D sample, TOI-1899 b is the only planet where $\porb$ is sufficiently large to keep a ratio with $\prot$ that allows the excitation of IWs. Regarding the rotation of the host star TOI-1899, for all the adopted $\prot$ the stellar rotation was spun down, and for those $\prot$ that were greater than $\sim10$ d, stellar rotation decreased to significantly lower values than the expected for M dwarfs, showing a difference between expected and calculated final rotation periods of more than 100\% for the most extreme cases. This probably means that there is a limit on how small the stellar rotation periods can initially be assumed, and that for $\prot\gtrsim10$ d tidal evolution is sufficiently slow that there is negligible tidal evolution within the age of the Universe for most parameters explored.

\begin{figure}
    \hspace{-0.2cm}
    \includegraphics[scale=0.69]{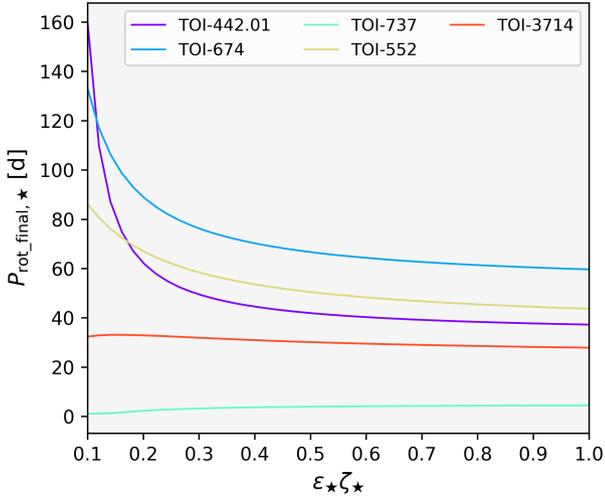}
    \caption{Final stellar rotation periods as a function of $\eps\gyrs$, for the planets in subsection \ref{sec:known} with known initial rotation periods (see Table \ref{tab:known}).}
    \label{fig:rotation_eps}
\end{figure}

For small $\prot$, the stellar rotation period of TOI-1899 also decreases significantly but within expected values of the rotation of M dwarfs. For instance, from the tidal evolution of $\prot=4.95$ d ($\tdecay=15.92$ Gyr) the final stellar rotation period was 13.94 d, rotating $\sim96\%$ faster than what is expected (615.5 d) from rotation-age relationships found empirically for M-dwarf hosts  \citep{Engle2018,Popinchalk2021} and similarly for other solar-type dwarfs \citep{Mamajek2008,Garcia2014}. This could indicate the effect that the spiral-in of planetary companions has on the rotation of their host star, showing that the planetary orbit transfers angular momentum that slows down the rate at which magnetic braking occurs and stellar rotation decreases. In general, the rapidly rotating scenarios of TOI-1899 b have a much stronger interaction with the host star since for IWs it is true that $\Qs\sim P_\mathrm{rot,\star}^2$, so more rapid rotation implies more rapid tidal evolution due to IWs.

TOI-1899 b undergoes orbital decay when $3.0\lesssim\prot\lesssim8.3$ d. For $\prot\gtrsim8.3$ d, the planet only reaches an asymptotic semi-major axis. This is due to the large migration time-scales involved in this system, as a direct consequence of the long initial orbital period when compared to other gas giants in this subgroup. As shown in Fig. \ref{fig:toi1899} for TOI-1899 b, $\tcircular$ is of the order of Gyr for this planet, thus meaning that circularization is achieved when the stellar rotational angular momentum has already been damped to a very low value due to the stellar wind's activity. This produces an inefficient transfer of angular momentum from the planetary orbit to the stellar spin (and vice versa). As a result, such exchange is not enough to shrink the orbit of the planet: in other words, orbital decay is suppressed.

While $\tcircular$ does not show a dependency on $\prot$ for any other planet in this subgroup (i.e. all the lines superpose when plotting their eccentricity damping), the circularization of TOI-1899 b depends on $\prot$ as shown by the lower sub-panel in Fig. \ref{fig:toi1899}. This is merely an effect of being located farther from its host star: the farther the planet is located the more time it has to tidally interact with the host star and thereby follow different evolutionary `paths'.

\subsection{Known stellar rotation periods}
\label{sec:known}

All the systems in this subgroup have a known $\prot$ (see Table \ref{tab:known}), so we use those values to study the effect of $\eps$ and $\gyrs$ (through $\eps\gyrs$) since these parameters could also affect the tidal evolution of compact star-planet systems \citep{Alvarado2021}. Since all of these stars have $0.42\leq\Mstar\leq0.6$, we set $0.1\leq\eps\gyrs\leq1.0$ (see e.g. \citealt{Gu2003,Dobs2004}). All the results for $\tdecay$ and $\tcircular$ of this subgroup are reported in Table \ref{tab:known}.  In general, it is worth noting that for all of the planets in this subgroup circularization occurs in time-scales of Myr. Moreover, $\tcircular$ is the same regardless of $\eps\gyrs$ due mainly to the very short orbital periods studied here, similar to the gas giants in Section \ref{sec:unknown} (except for TOI-1899 b).

\setlength{\tabcolsep}{3pt}
\begin{table}
\centering
    \begin{tabular}{c|c|c|c|c|c}
         \hline\hline planet & $e$ & $\porb$ [d] & $\prot$ [d] & $\tdecay$ [Gyr] & $\tcircular$ [Myr] \\ \hline
         TOI-737 b & 0.044 & 1.73185606 & 4.745422 & 0.54 -- 0.89 & 6.17 \\
         TOI-442.01 & 0.04 & 4.052037 & 33 & 1.72 -- 22.43 & 21.67\\
         TOI-674 b & 0.1 & 1.977143 & 52 & 1.51 -- 10.36 & 0.68 \\
         TOI-3714 b & 0.03 & 2.154849 & 23.4 & 1.92 -- 5.58 & 6.41\\
         TOI-552 b & 0.064 & 2.7886556 & 35.0435 & 6.73 -- 29.5 & 16.25\\
         \hline
    \end{tabular}
    \caption{Gas giants in subsection \ref{sec:known}. Here, $\tdecay$ and $\tcircular$ are from this work, whereas $e$,  $\porb$, and $\prot$ are extracted from the references in Section \ref{sec:intro}.}
    \label{tab:known}
    \vspace{-0.2cm}
\end{table}

Tidal torques act on the whole convective envelope of the star and the planet, but as explained in \citet{Dobs2004} the subsequent transfer of angular momentum could be a function of the stellar mass fraction that participates dynamically in the exchange of angular momentum as well as of spectral type \citep*{Pinsonneault2001}. In order to mathematically analyse how the stellar spin and planetary orbit are affected if only a fraction of the stellar mass participated in the transfer of angular momentum, we artificially add $\eps$ and $\epp$ to the expressions of the stellar and planetary moments of inertia (i.e. equations \ref{eq:moment_s} and \ref{eq:moment_p}, respectively), similar to equations 25 and 26 in \citet{Dobs2004}.

Such quantities are numerical variations which produce different solutions for the differential equations that describe the tidal evolution of each system. By numerically assuming that only a part of the stellar envelope exchanges rotational angular momentum with the planetary orbit, we seek to analyse the effect that the stellar mass has on stellar tidal dissipation and orbital evolution of close-in gas giants. For the stars in the GGM-D sample, we are assuming that both radiative core and convective envelope dissipate energy, so $\eps$ will describe the fraction of the total stellar mass that participates dynamically. However, this is just a numerical experiment and over stellar evolutionary time-scales such quantities might differ from the assumed values in this work. The aim of this analysis is just to see if any variations in their mass may overestimate or underestimate the time-scales of tidal evolution for the internal mechanisms of angular momentum transport considered in this work.

As depicted in Figs \ref{fig:toi442}, \ref{fig:toi552}, \ref{fig:toi674}, \ref{fig:toi3714}, and \ref{fig:toi737} orbital decay is impending and happens for any value $\eps\gyrs$. For less massive stellar envelopes (i.e. low $\eps\gyrs$), the evolution of the planetary orbit takes longer to reach the Roche limit in each system, being an effect of the long $\prot$ of TOI-442.01, TOI-552, TOI-674, and TOI-3714 as both effects add up: 1) having a small portion of the stellar envelope transferring angular momentum to the planetary orbit, and 2) the stellar spin being so slow that such a transfer is inefficient. It is worth noting that high values of the product $\eps\gyrs$ correspond to most of the cases studied in the literature and that migration time-scales for low $\eps\gyrs$ should be taken with a grain of salt due to the high uncertainty in the quantity $\eps$. For any planet in this subgroup with known $\prot$, the orbital migration time-scales between extreme values of $\eps\gyrs$ differ in one order of magnitude, which is expected as $\tdecay$ and/or $\tcircular$ are not determined from first principles and could be over- or under-estimated.

The idea that only a portion of the star might exchange angular momentum with its companion is worthwhile to be pursued in order to constrain the orbital evolution time-scales of the planet. These variations in the stellar moment of inertia are directly related to the coupling between radiative zones and convective regions of a star, exchanging angular momentum on a given time-scale, and introduced in other works to account for the removal of differential rotation over stellar (and tidal) evolutionary time-scales due to magneto-hydrodynamic turbulence and gravity waves \citep{Barker2009,Winn2010}. The time-scales that we found here for small $\eps\gyrs$ are over-estimated with respect to constraints calculated in some observational efforts \citep{Gallet2013}, posing a problem when using core-envelope decoupling models, although these time-scales are still observationally poorly constrained.

In Fig. \ref{fig:rotation_eps}, we show the final stellar rotation period as a function of $\eps\gyrs$. We can see that small $\eps\gyrs$ led to longer final stellar rotation periods, where TOI-737 and TOI-442.01 had the shortest (i.e. $\sim2$ d) and longest (i.e. $\sim160$ d) values, respectively. As $\eps\gyrs$ gets larger, the transfer of angular momentum from the planetary orbit to the stellar spin is more significant and this balances out (at least momentarily) the loss of angular momentum due to magnetic braking. This slows down the decreasing rate of the stellar rotational rate and produce shorter final rotation periods. For middle values of $\eps\gyrs$, we can see that from $\eps\gyrs\simeq0.5$ such a product (and hence parameters) does not significantly affect the stellar rotation rate and that the change is almost linear with a very moderate gradient. Interestingly, for large $\eps\gyrs$ the stellar rotation period decreased by 30\% (TOI-737), 28\% (TOI-3714), 28\% (TOI-552), 21\% (TOI-442.01), and 20\% (TOI-674) in comparison to more than 90\% for small $\eps\gyrs$.

\section{Discussion}
\label{sec:concl}

This is the first intensive study of tidally-induced migration of the only discovered population of gas giants orbiting M-dwarf hosts (see Fig. \ref{fig:distri}). To improve models that describe stellar and planetary evolution, it is important that we understand how such systems will evolve and analyse the entwined relationships between their orbital and physical parameters. For example, by the end of core-accretion gas giants could be inflated and this might modify the thermal and dynamical evolution of hot Jupiters migrating towards their host star \citep{Rozner2022,Glanz2022}, so similar effects might be relevant for gas giants around M dwarfs too. Still, how these gas giants reached their current position needs a backward integration approach, which will allow tracking their initial position before tidal evolution started: the tidal history of such planets dictates that they might have also evolved possibly due to tidal effects with their host star. This would complete the picture of their orbital migration, but we will conduct future research on this and for now, it is out of the scope of this work.

We focus on the forward tidal evolution of all the gas giants orbiting M dwarfs discovered/confirmed by TESS (which we called the GGM-D sample). The torques acting on both the star and the planet resulting from tidal interactions can have some effects on the natural evolution of their rotation. For these systems, we found that the longer the orbital period the more insignificant the effect on the stellar spin, but if planets are located closer this might have consequences for stellar age estimations: M dwarfs with a massive gas giant companion can undergo a significant transfer of angular momentum, changing the stellar spin from expected values and leading to a biased over- or under-estimation of their age when using, for instance, analysis such as gyrochronology (see e.g. \citealt{Gallet2020}). This can affect the estimation of other system properties (e.g. metallicity), particularly if such properties rely on current stellar and planetary models.

We have analysed the tidally-induced orbital migration of TESS gas giants around M dwarfs, including tidal mechanisms that are thought to have the most significant contributions to the dissipation of energy in compact planetary systems. Such mechanisms are the excitation of inertial waves (IWs) in stellar convective envelopes and internal gravity waves (IGWs) in radiative zones. IWs were more important for rapidly rotating host stars when compared to the mean motion of their planetary companion, and we found that IWs drove the evolution of systems such as TOI-737 b and TOI-1899 b, as both of them had a tidal frequency comparable to the spin rate of the star. For the rest of the systems, IWs were not significant since the rotation of the stars was not enough to excite this type of waves in the stellar convective envelopes. Instead, energy dissipation due to the damping of IGWs in radiative zones was the dominant mechanism driving the orbital evolution of the planets, having a very notable effect for those systems where planets were located at very short-period orbits.

Regarding the above, for very small $\eps\gyrs$ (see Section \ref{sec:known}) we found that only TOI-737 was subject to a significant spin-up and decreased its rotation period approximately to 80\% of its original value while the planet was on its way to the Roche limit. All of the other planets in the GGM-D sample did not produce a prominent acceleration of the stellar rotation, at least not until they approached their Roche limit. In general, once planets cross the Roche limit the stellar rotation accelerates more significantly, but still, a fast planetary orbital decay (and hence a short orbital period) is the key factor that will determine whether a star will be spun up or spun down. In this regard, by constraining $\frac{\Der\Os}{\Der t}$, we might be able to extrapolate backwards and find how much the stellar rotational rate has increased or decreased due to tidal interactions. This would allow us to improve age estimations and recalculate other stellar properties, and it is being a matter of study for our next research work.

There are still many unknowns when it comes to energy dissipation in stars and planets: how is the heat transferred from the core to the envelope? what role do boundary layers play in the overall heat dissipation? how is their physical deformation affected by the system's angular momentum exchange (and vice versa)? Addressing these questions will allow us to explain the orbital configuration of different types of systems and shed some light on the physical properties of both host stars and planetary companions. Studying specific systems such as those analysed here will allow us to know more about stellar evolution and planetary formation, find out their possible past, and constrain their fate moving forward.

\vspace{-0.4cm}
\section*{Acknowledgements}

The author thanks the anonymous referee for a very valuable review that significantly improved the quality and context of this research. Also, thanks to Tianjun Gan and Mor Rozner for valuable comments on this research. JAA-M is funded by the International Macquarie University Research Excellence Scholarship (`iMQRES'). This research has made use of the NASA's Astrophysics Data System (ADS) and NASA Exoplanet Archive, which is operated by the California Institute of Technology, under contract with the National Aeronautics and Space Administration under the Exoplanet Exploration Program.


\section*{Data Availability}
The data underlying this article will be shared on reasonable request to the corresponding author.




\begin{thebibliography}{}
\makeatletter
\relax
\def\mn@urlcharsother{\let\do\@makeother \do\$\do\&\do\#\do\^\do\_\do\%\do\~}
\def\mn@doi{\begingroup\mn@urlcharsother \@ifnextchar [ {\mn@doi@}
  {\mn@doi@[]}}
\def\mn@doi@[#1]#2{\def\@tempa{#1}\ifx\@tempa\@empty \href
  {http://dx.doi.org/#2} {doi:#2}\else \href {http://dx.doi.org/#2} {#1}\fi
  \endgroup}
\def\mn@eprint#1#2{\mn@eprint@#1:#2::\@nil}
\def\mn@eprint@arXiv#1{\href {http://arxiv.org/abs/#1} {{\tt arXiv:#1}}}
\def\mn@eprint@dblp#1{\href {http://dblp.uni-trier.de/rec/bibtex/#1.xml}
  {dblp:#1}}
\def\mn@eprint@#1:#2:#3:#4\@nil{\def\@tempa {#1}\def\@tempb {#2}\def\@tempc
  {#3}\ifx \@tempc \@empty \let \@tempc \@tempb \let \@tempb \@tempa \fi \ifx
  \@tempb \@empty \def\@tempb {arXiv}\fi \@ifundefined
  {mn@eprint@\@tempb}{\@tempb:\@tempc}{\expandafter \expandafter \csname
  mn@eprint@\@tempb\endcsname \expandafter{\@tempc}}}

\bibitem[\protect\citeauthoryear{{Alexander}}{{Alexander}}{1973}]{Alexander1973}
{Alexander} M.~E.,  1973, \mn@doi [\apss] {10.1007/BF00645172}, \href
  {https://ui.adsabs.harvard.edu/abs/1973Ap&SS..23..459A} {23, 459}

\bibitem[\protect\citeauthoryear{{Alvarado-Montes} \&
  {Garc{\'\i}a-Carmona}}{{Alvarado-Montes} \&
  {Garc{\'\i}a-Carmona}}{2019}]{Alvarado2019}
{Alvarado-Montes} J.~A.,  {Garc{\'\i}a-Carmona} C.,  2019, \mn@doi [\mnras]
  {10.1093/mnras/stz1081}, \href
  {https://ui.adsabs.harvard.edu/abs/2019MNRAS.486.3963A} {486, 3963}

\bibitem[\protect\citeauthoryear{{Alvarado-Montes}, {Zuluaga}  \&
  {Sucerquia}}{{Alvarado-Montes} et~al.}{2017}]{Alvarado2017}
{Alvarado-Montes} J.~A.,  {Zuluaga} J.~I.,   {Sucerquia} M.,  2017, \mn@doi
  [\mnras] {10.1093/mnras/stx1745}, \href
  {http://adsabs.harvard.edu/abs/2017MNRAS.471.3019A} {471, 3019}

\bibitem[\protect\citeauthoryear{{Alvarado-Montes}, {Sucerquia},
  {Garc{\'\i}a-Carmona}, {Zuluaga}, {Spitler}  \& {Schwab}}{{Alvarado-Montes}
  et~al.}{2021}]{Alvarado2021}
{Alvarado-Montes} J.~A.,  {Sucerquia} M.,  {Garc{\'\i}a-Carmona} C.,  {Zuluaga}
  J.~I.,  {Spitler} L.,   {Schwab} C.,  2021, \mn@doi [\mnras]
  {10.1093/mnras/stab1081}, \href
  {https://ui.adsabs.harvard.edu/abs/2021MNRAS.506.2247A} {506, 2247}

\bibitem[\protect\citeauthoryear{{Alves}, {Do Nascimento}  \& {de
  Medeiros}}{{Alves} et~al.}{2010}]{Alves2010}
{Alves} S.,  {Do Nascimento} J.~D. J.,   {de Medeiros} J.~R.,  2010, \mn@doi
  [\mnras] {10.1111/j.1365-2966.2010.17243.x}, \href
  {https://ui.adsabs.harvard.edu/abs/2010MNRAS.408.1770A} {408, 1770}

\bibitem[\protect\citeauthoryear{{Barker}}{{Barker}}{2011}]{Barker2011}
{Barker} A.~J.,  2011, \mn@doi [\mnras] {10.1111/j.1365-2966.2011.18468.x},
  \href {https://ui.adsabs.harvard.edu/abs/2011MNRAS.414.1365B} {414, 1365}

\bibitem[\protect\citeauthoryear{{Barker}}{{Barker}}{2020}]{Barker2020}
{Barker} A.~J.,  2020, \mn@doi [\mnras] {10.1093/mnras/staa2405}, \href
  {https://ui.adsabs.harvard.edu/abs/2020MNRAS.498.2270B} {498, 2270}

\bibitem[\protect\citeauthoryear{{Barker} \& {Ogilvie}}{{Barker} \&
  {Ogilvie}}{2009}]{Barker2009}
{Barker} A.~J.,  {Ogilvie} G.~I.,  2009, \mn@doi [\mnras]
  {10.1111/j.1365-2966.2009.14694.x}, \href
  {https://ui.adsabs.harvard.edu/\#abs/2009MNRAS.395.2268B} {395, 2268}

\bibitem[\protect\citeauthoryear{{Barker} \& {Ogilvie}}{{Barker} \&
  {Ogilvie}}{2010}]{Barker2010}
{Barker} A.~J.,  {Ogilvie} G.~I.,  2010, \mn@doi [\mnras]
  {10.1111/j.1365-2966.2010.16400.x}, \href
  {https://ui.adsabs.harvard.edu/abs/2010MNRAS.404.1849B} {404, 1849}

\bibitem[\protect\citeauthoryear{{Bolton} et~al.,}{{Bolton}
  et~al.}{2017}]{Bolton2017}
{Bolton} S.~J.,  et~al., 2017, \mn@doi [\ssr] {10.1007/s11214-017-0429-6},
  \href {https://ui.adsabs.harvard.edu/abs/2017SSRv..213....5B} {213, 5}

\bibitem[\protect\citeauthoryear{{Brown}, {Collier Cameron}, {Hall}, {Hebb}  \&
  {Smalley}}{{Brown} et~al.}{2011}]{Brown2011}
{Brown} D.~J.~A.,  {Collier Cameron} A.,  {Hall} C.,  {Hebb} L.,   {Smalley}
  B.,  2011, \mn@doi [\mnras] {10.1111/j.1365-2966.2011.18729.x}, \href
  {https://ui.adsabs.harvard.edu/abs/2011MNRAS.415..605B} {415, 605}

\bibitem[\protect\citeauthoryear{{Ca{\~n}as} et~al.,}{{Ca{\~n}as}
  et~al.}{2020}]{Canas2020}
{Ca{\~n}as} C.~I.,  et~al., 2020, \mn@doi [\aj] {10.3847/1538-3881/abac67},
  \href {https://ui.adsabs.harvard.edu/abs/2020AJ....160..147C} {160, 147}

\bibitem[\protect\citeauthoryear{{Ca{\~n}as} et~al.,}{{Ca{\~n}as}
  et~al.}{2022}]{Canas2022}
{Ca{\~n}as} C.~I.,  et~al., 2022, \mn@doi [\aj] {10.3847/1538-3881/ac7804},
  \href {https://ui.adsabs.harvard.edu/abs/2022AJ....164...50C} {164, 50}

\bibitem[\protect\citeauthoryear{{Collier Cameron} \& {Li}}{{Collier Cameron}
  \& {Li}}{1994}]{Cameron1994}
{Collier Cameron} A.,  {Li} J.,  1994, \mn@doi [\mnras]
  {10.1093/mnras/269.4.1099}, \href
  {https://ui.adsabs.harvard.edu/abs/1994MNRAS.269.1099C} {269, 1099}

\bibitem[\protect\citeauthoryear{{Darwin}}{{Darwin}}{1879}]{Darwin1879}
{Darwin} G.~H.,  1879, Philosophical Transactions of the Royal Society of
  London Series I, \href
  {https://ui.adsabs.harvard.edu/abs/1879RSPT..170....1D} {170, 1}

\bibitem[\protect\citeauthoryear{{Debras} \& {Chabrier}}{{Debras} \&
  {Chabrier}}{2019}]{Debras2019}
{Debras} F.,  {Chabrier} G.,  2019, \mn@doi [\apj] {10.3847/1538-4357/aaff65},
  \href {https://ui.adsabs.harvard.edu/abs/2019ApJ...872..100D} {872, 100}

\bibitem[\protect\citeauthoryear{{Dobbs-Dixon}, {Lin}  \&
  {Mardling}}{{Dobbs-Dixon} et~al.}{2004}]{Dobs2004}
{Dobbs-Dixon} I.,  {Lin} D.~N.~C.,   {Mardling} R.~A.,  2004, \mn@doi [\apj]
  {10.1086/421510}, \href
  {https://ui.adsabs.harvard.edu/\#abs/2004ApJ...610..464D} {610, 464}

\bibitem[\protect\citeauthoryear{{Dreizler} et~al.,}{{Dreizler}
  et~al.}{2020}]{Dreizler2020}
{Dreizler} S.,  et~al., 2020, \mn@doi [\aap] {10.1051/0004-6361/202038016},
  \href {https://ui.adsabs.harvard.edu/abs/2020A&A...644A.127D} {644, A127}

\bibitem[\protect\citeauthoryear{{Duguid}, {Barker}  \& {Jones}}{{Duguid}
  et~al.}{2020a}]{Duguid2020a}
{Duguid} C.~D.,  {Barker} A.~J.,   {Jones} C.~A.,  2020a, \mn@doi [\mnras]
  {10.1093/mnras/stz2899}, \href
  {https://ui.adsabs.harvard.edu/abs/2020MNRAS.491..923D} {491, 923}

\bibitem[\protect\citeauthoryear{{Duguid}, {Barker}  \& {Jones}}{{Duguid}
  et~al.}{2020b}]{Duguid2020b}
{Duguid} C.~D.,  {Barker} A.~J.,   {Jones} C.~A.,  2020b, \mn@doi [\mnras]
  {10.1093/mnras/staa2216}, \href
  {https://ui.adsabs.harvard.edu/abs/2020MNRAS.497.3400D} {497, 3400}

\bibitem[\protect\citeauthoryear{{Efroimsky}}{{Efroimsky}}{2012}]{Efroimsky2012}
{Efroimsky} M.,  2012, \mn@doi [\apj] {10.1088/0004-637X/746/2/150}, \href
  {https://ui.adsabs.harvard.edu/\#abs/2012ApJ...746..150E} {746, 150}

\bibitem[\protect\citeauthoryear{{Engle} \& {Guinan}}{{Engle} \&
  {Guinan}}{2018}]{Engle2018}
{Engle} S.~G.,  {Guinan} E.~F.,  2018, \mn@doi [Research Notes of the American
  Astronomical Society] {10.3847/2515-5172/aab1f8}, \href
  {https://ui.adsabs.harvard.edu/abs/2018RNAAS...2...34E} {2, 34}

\bibitem[\protect\citeauthoryear{{Finley} \& {Matt}}{{Finley} \&
  {Matt}}{2017}]{Finley2017}
{Finley} A.~J.,  {Matt} S.~P.,  2017, \mn@doi [\apj]
  {10.3847/1538-4357/aa7fb9}, \href
  {https://ui.adsabs.harvard.edu/abs/2017ApJ...845...46F} {845, 46}

\bibitem[\protect\citeauthoryear{{Gallet}}{{Gallet}}{2020}]{Gallet2020}
{Gallet} F.,  2020, \mn@doi [\aap] {10.1051/0004-6361/202038058}, \href
  {https://ui.adsabs.harvard.edu/abs/2020A&A...641A..38G} {641, A38}

\bibitem[\protect\citeauthoryear{{Gallet} \& {Bouvier}}{{Gallet} \&
  {Bouvier}}{2013}]{Gallet2013}
{Gallet} F.,  {Bouvier} J.,  2013, \mn@doi [\aap]
  {10.1051/0004-6361/201321302}, \href
  {https://ui.adsabs.harvard.edu/abs/2013A&A...556A..36G} {556, A36}

\bibitem[\protect\citeauthoryear{{Gallet}, {Bolmont}, {Mathis}, {Charbonnel}
  \& {Amard}}{{Gallet} et~al.}{2017}]{Gallet2017}
{Gallet} F.,  {Bolmont} E.,  {Mathis} S.,  {Charbonnel} C.,   {Amard} L.,
  2017, \mn@doi [\aap] {10.1051/0004-6361/201730661}, \href
  {https://ui.adsabs.harvard.edu/abs/2017A&A...604A.112G} {604, A112}

\bibitem[\protect\citeauthoryear{{Gan} et~al.,}{{Gan} et~al.}{2022}]{Gan2022}
{Gan} T.,  et~al., 2022, \mn@doi [\mnras] {10.1093/mnras/stab3708}, \href
  {https://ui.adsabs.harvard.edu/abs/2022MNRAS.511...83G} {511, 83}

\bibitem[\protect\citeauthoryear{{Garc{\'\i}a} et~al.,}{{Garc{\'\i}a}
  et~al.}{2014}]{Garcia2014}
{Garc{\'\i}a} R.~A.,  et~al., 2014, \mn@doi [\aap]
  {10.1051/0004-6361/201423888}, \href
  {https://ui.adsabs.harvard.edu/abs/2014A&A...572A..34G} {572, A34}

\bibitem[\protect\citeauthoryear{{Glanz}, {Rozner}, {Perets}  \&
  {Grishin}}{{Glanz} et~al.}{2022}]{Glanz2022}
{Glanz} H.,  {Rozner} M.,  {Perets} H.~B.,   {Grishin} E.,  2022, \mn@doi
  [\apj] {10.3847/1538-4357/ac6807}, \href
  {https://ui.adsabs.harvard.edu/abs/2022ApJ...931...11G} {931, 11}

\bibitem[\protect\citeauthoryear{{Goldreich} \& {Nicholson}}{{Goldreich} \&
  {Nicholson}}{1977}]{Goldreich1977}
{Goldreich} P.,  {Nicholson} P.~D.,  1977, \mn@doi [\icarus]
  {10.1016/0019-1035(77)90163-4}, \href
  {https://ui.adsabs.harvard.edu/\#abs/1977Icar...30..301G} {30, 301}

\bibitem[\protect\citeauthoryear{{Goldreich} \& {Soter}}{{Goldreich} \&
  {Soter}}{1966}]{Goldreich1966}
{Goldreich} P.,  {Soter} S.,  1966, \mn@doi [icarus]
  {10.1016/0019-1035(66)90051-0}, \href
  {https://ui.adsabs.harvard.edu/\#abs/1966Icar....5..375G} {5, 375}

\bibitem[\protect\citeauthoryear{{Gu}, {Lin}  \& {Bodenheimer}}{{Gu}
  et~al.}{2003}]{Gu2003}
{Gu} P.-G.,  {Lin} D. N.~C.,   {Bodenheimer} P.~H.,  2003, \mn@doi [\apj]
  {10.1086/373920}, \href
  {https://ui.adsabs.harvard.edu/\#abs/2003ApJ...588..509G} {588, 509}

\bibitem[\protect\citeauthoryear{{Gu}, {Bodenheimer}  \& {Lin}}{{Gu}
  et~al.}{2004}]{Gu2004}
{Gu} P.-G.,  {Bodenheimer} P.~H.,   {Lin} D. N.~C.,  2004, \mn@doi [\apj]
  {10.1086/420867}, \href
  {https://ui.adsabs.harvard.edu/\#abs/2004ApJ...608.1076G} {608, 1076}

\bibitem[\protect\citeauthoryear{{Guenel}, {Mathis}  \& {Remus}}{{Guenel}
  et~al.}{2014}]{Guenel2014}
{Guenel} M.,  {Mathis} S.,   {Remus} F.,  2014, \mn@doi [\aap]
  {10.1051/0004-6361/201424010}, \href
  {http://adsabs.harvard.edu/abs/2014A%26A...566L...9G} {566, L9}

\bibitem[\protect\citeauthoryear{{Guillochon}, {Ramirez-Ruiz}  \&
  {Lin}}{{Guillochon} et~al.}{2011}]{Guillochon2011}
{Guillochon} J.,  {Ramirez-Ruiz} E.,   {Lin} D.,  2011, \mn@doi [\apj]
  {10.1088/0004-637X/732/2/74}, \href
  {https://ui.adsabs.harvard.edu/\#abs/2011ApJ...732...74G} {732, 74}

\bibitem[\protect\citeauthoryear{{Hansen}}{{Hansen}}{2010}]{Hansen2010}
{Hansen} B. M.~S.,  2010, \mn@doi [\apj] {10.1088/0004-637X/723/1/285}, \href
  {https://ui.adsabs.harvard.edu/\#abs/2010ApJ...723..285H} {723, 285}

\bibitem[\protect\citeauthoryear{{Hut}}{{Hut}}{1980}]{Hut1980}
{Hut} P.,  1980, \aap, \href
  {https://ui.adsabs.harvard.edu/abs/1980A&A....92..167H} {92, 167}

\bibitem[\protect\citeauthoryear{{Hut}}{{Hut}}{1981}]{Hut1981}
{Hut} P.,  1981, \aap, \href
  {https://ui.adsabs.harvard.edu/\#abs/1981A&A....99..126H} {99, 126}

\bibitem[\protect\citeauthoryear{{Ivanov}, {Papaloizou}  \& {Chernov}}{{Ivanov}
  et~al.}{2013}]{Ivanov2013}
{Ivanov} P.~B.,  {Papaloizou} J.~C.~B.,   {Chernov} S.~V.,  2013, \mn@doi
  [\mnras] {10.1093/mnras/stt595}, \href
  {http://adsabs.harvard.edu/abs/2013MNRAS.432.2339I} {432, 2339}

\bibitem[\protect\citeauthoryear{{Jackson}, {Greenberg}  \& {Barnes}}{{Jackson}
  et~al.}{2008}]{Jackson2008a}
{Jackson} B.,  {Greenberg} R.,   {Barnes} R.,  2008, \mn@doi [\apj]
  {10.1086/529187}, \href
  {https://ui.adsabs.harvard.edu/\#abs/2008ApJ...678.1396J} {678, 1396}

\bibitem[\protect\citeauthoryear{{Jord{\'a}n} et~al.,}{{Jord{\'a}n}
  et~al.}{2022}]{Jordan2022}
{Jord{\'a}n} A.,  et~al., 2022, \mn@doi [\aj] {10.3847/1538-3881/ac4a77}, \href
  {https://ui.adsabs.harvard.edu/abs/2022AJ....163..125J} {163, 125}

\bibitem[\protect\citeauthoryear{{Kanodia} et~al.,}{{Kanodia}
  et~al.}{2020}]{Kanodia2020}
{Kanodia} S.,  et~al., 2020, \mn@doi [\apj] {10.3847/1538-4357/aba0a2}, \href
  {https://ui.adsabs.harvard.edu/abs/2020ApJ...899...29K} {899, 29}

\bibitem[\protect\citeauthoryear{{Kanodia} et~al.,}{{Kanodia}
  et~al.}{2021}]{Kanodia2021}
{Kanodia} S.,  et~al., 2021, \mn@doi [\aj] {10.3847/1538-3881/ac1940}, \href
  {https://ui.adsabs.harvard.edu/abs/2021AJ....162..135K} {162, 135}

\bibitem[\protect\citeauthoryear{{Kanodia} et~al.,}{{Kanodia}
  et~al.}{2022}]{Kanodia2022}
{Kanodia} S.,  et~al., 2022, \mn@doi [\aj] {10.3847/1538-3881/ac7c20}, \href
  {https://ui.adsabs.harvard.edu/abs/2022AJ....164...81K} {164, 81}

\bibitem[\protect\citeauthoryear{{Lainey}, {Arlot}, {Karatekin}  \& {van
  Hoolst}}{{Lainey} et~al.}{2009}]{Lainey2009}
{Lainey} V.,  {Arlot} J.-E.,  {Karatekin} {\"O}.,   {van Hoolst} T.,  2009,
  \mn@doi [\nat] {10.1038/nature08108}, \href
  {https://ui.adsabs.harvard.edu/\#abs/2009Natur.459..957L} {459, 957}

\bibitem[\protect\citeauthoryear{{Lainey} et~al.,}{{Lainey}
  et~al.}{2012}]{Lainey2012}
{Lainey} V.,  et~al., 2012, \mn@doi [\apj] {10.1088/0004-637X/752/1/14}, \href
  {http://adsabs.harvard.edu/abs/2012ApJ...752...14L} {752, 14}

\bibitem[\protect\citeauthoryear{{Lanza}}{{Lanza}}{2010}]{Lanza2010}
{Lanza} A.~F.,  2010, \mn@doi [\aap] {10.1051/0004-6361/200912789}, \href
  {https://ui.adsabs.harvard.edu/abs/2010A&A...512A..77L} {512, A77}

\bibitem[\protect\citeauthoryear{{Leconte}, {Chabrier}, {Baraffe}  \&
  {Levrard}}{{Leconte} et~al.}{2010}]{Leconte2010}
{Leconte} J.,  {Chabrier} G.,  {Baraffe} I.,   {Levrard} B.,  2010, \mn@doi
  [\aap] {10.1051/0004-6361/201014337}, \href
  {https://ui.adsabs.harvard.edu/abs/2010A&A...516A..64L} {516, A64}

\bibitem[\protect\citeauthoryear{{Love}}{{Love}}{1927}]{Love1927}
{Love} A.~E.~H.,  1927, {A Treatise on the Mathematical Theory of Elasticity}.
~ Vol. 4, Cambridge University Press

\bibitem[\protect\citeauthoryear{{Mamajek} \& {Hillenbrand}}{{Mamajek} \&
  {Hillenbrand}}{2008}]{Mamajek2008}
{Mamajek} E.~E.,  {Hillenbrand} L.~A.,  2008, \mn@doi [\apj] {10.1086/591785},
  \href {https://ui.adsabs.harvard.edu/abs/2008ApJ...687.1264M} {687, 1264}

\bibitem[\protect\citeauthoryear{{Mardling} \& {Lin}}{{Mardling} \&
  {Lin}}{2002}]{Mardling2002}
{Mardling} R.~A.,  {Lin} D.~N.~C.,  2002, \mn@doi [\apj] {10.1086/340752},
  \href {https://ui.adsabs.harvard.edu/\#abs/2002ApJ...573..829M} {573, 829}

\bibitem[\protect\citeauthoryear{{Mathis}}{{Mathis}}{2015}]{Mathis2015b}
{Mathis} S.,  2015, \mn@doi [\aap] {10.1051/0004-6361/201526472}, \href
  {http://adsabs.harvard.edu/abs/2015A\%26A...580L...3M} {580, L3}

\bibitem[\protect\citeauthoryear{{M{\"u}ller}, {Helled}  \&
  {Cumming}}{{M{\"u}ller} et~al.}{2020}]{Muller2020}
{M{\"u}ller} S.,  {Helled} R.,   {Cumming} A.,  2020, \mn@doi [\aap]
  {10.1051/0004-6361/201937376}, \href
  {https://ui.adsabs.harvard.edu/abs/2020A&A...638A.121M} {638, A121}

\bibitem[\protect\citeauthoryear{{Murgas} et~al.,}{{Murgas}
  et~al.}{2021}]{Murgas2021}
{Murgas} F.,  et~al., 2021, \mn@doi [\aap] {10.1051/0004-6361/202140718}, \href
  {https://ui.adsabs.harvard.edu/abs/2021A&A...653A..60M} {653, A60}

\bibitem[\protect\citeauthoryear{{Ogilvie}}{{Ogilvie}}{2013}]{Ogilvie2013}
{Ogilvie} G.~I.,  2013, \mn@doi [\mnras] {10.1093/mnras/sts362}, \href
  {http://adsabs.harvard.edu/abs/2013MNRAS.429..613O} {429, 613}

\bibitem[\protect\citeauthoryear{{Ogilvie} \& {Lin}}{{Ogilvie} \&
  {Lin}}{2004}]{Ogilvie2004}
{Ogilvie} G.~I.,  {Lin} D.~N.~C.,  2004, \mn@doi [\apj] {10.1086/421454}, \href
  {https://ui.adsabs.harvard.edu/\#abs/2004ApJ...610..477O} {610, 477}

\bibitem[\protect\citeauthoryear{{Ogilvie} \& {Lin}}{{Ogilvie} \&
  {Lin}}{2007}]{Ogilvie2007}
{Ogilvie} G.~I.,  {Lin} D.~N.~C.,  2007, \mn@doi [\apj] {10.1086/515435}, \href
  {http://adsabs.harvard.edu/abs/2007ApJ...661.1180O} {661, 1180}

\bibitem[\protect\citeauthoryear{{Papaloizou} \& {Savonije}}{{Papaloizou} \&
  {Savonije}}{1997}]{Papaloizou1997}
{Papaloizou} J.~C.~B.,  {Savonije} G.~J.,  1997, \mn@doi [\mnras]
  {10.1093/mnras/291.4.651}, \href
  {https://ui.adsabs.harvard.edu/abs/1997MNRAS.291..651P} {291, 651}

\bibitem[\protect\citeauthoryear{{Penev}, {Barranco}  \& {Sasselov}}{{Penev}
  et~al.}{2009}]{Penev2009}
{Penev} K.,  {Barranco} J.,   {Sasselov} D.,  2009, \mn@doi [\apj]
  {10.1088/0004-637X/705/1/285}, \href
  {https://ui.adsabs.harvard.edu/abs/2009ApJ...705..285P} {705, 285}

\bibitem[\protect\citeauthoryear{{Pinsonneault}, {DePoy}  \&
  {Coffee}}{{Pinsonneault} et~al.}{2001}]{Pinsonneault2001}
{Pinsonneault} M.~H.,  {DePoy} D.~L.,   {Coffee} M.,  2001, \mn@doi [\apjl]
  {10.1086/323531}, \href
  {https://ui.adsabs.harvard.edu/abs/2001ApJ...556L..59P} {556, L59}

\bibitem[\protect\citeauthoryear{{Popinchalk}, {Faherty}, {Kiman}, {Gagn{\'e}},
  {Curtis}, {Angus}, {Cruz}  \& {Rice}}{{Popinchalk}
  et~al.}{2021}]{Popinchalk2021}
{Popinchalk} M.,  {Faherty} J.~K.,  {Kiman} R.,  {Gagn{\'e}} J.,  {Curtis}
  J.~L.,  {Angus} R.,  {Cruz} K.~L.,   {Rice} E.~L.,  2021, \mn@doi [\apj]
  {10.3847/1538-4357/ac0444}, \href
  {https://ui.adsabs.harvard.edu/abs/2021ApJ...916...77P} {916, 77}

\bibitem[\protect\citeauthoryear{{R{\'e}ville}, {Brun}, {Matt}, {Strugarek}  \&
  {Pinto}}{{R{\'e}ville} et~al.}{2015}]{Reville2015a}
{R{\'e}ville} V.,  {Brun} A.~S.,  {Matt} S.~P.,  {Strugarek} A.,   {Pinto}
  R.~F.,  2015, \mn@doi [\apj] {10.1088/0004-637X/798/2/116}, \href
  {https://ui.adsabs.harvard.edu/abs/2015ApJ...798..116R} {798, 116}

\bibitem[\protect\citeauthoryear{{R{\'e}ville}, {Folsom}, {Strugarek}  \&
  {Brun}}{{R{\'e}ville} et~al.}{2016}]{Reville2016a}
{R{\'e}ville} V.,  {Folsom} C.~P.,  {Strugarek} A.,   {Brun} A.~S.,  2016,
  \mn@doi [\apj] {10.3847/0004-637X/832/2/145}, \href
  {https://ui.adsabs.harvard.edu/abs/2016ApJ...832..145R} {832, 145}

\bibitem[\protect\citeauthoryear{{Ricker} et~al.,}{{Ricker}
  et~al.}{2015}]{Ricker2015}
{Ricker} G.~R.,  et~al., 2015, \mn@doi [JATIS] {10.1117/1.JATIS.1.1.014003},
  \href {http://adsabs.harvard.edu/abs/2015JATIS...1a4003R} {1, 014003}

\bibitem[\protect\citeauthoryear{{Roche}}{{Roche}}{1849}]{Roche1849}
{Roche} E.,  1849, Acad. Montpellier, 1, 243

\bibitem[\protect\citeauthoryear{{Rozner}, {Glanz}, {Perets}  \&
  {Grishin}}{{Rozner} et~al.}{2022}]{Rozner2022}
{Rozner} M.,  {Glanz} H.,  {Perets} H.~B.,   {Grishin} E.,  2022, \mn@doi
  [\apj] {10.3847/1538-4357/ac6808}, \href
  {https://ui.adsabs.harvard.edu/abs/2022ApJ...931...10R} {931, 10}

\bibitem[\protect\citeauthoryear{{Skumanich}}{{Skumanich}}{1972}]{Skumanich1972}
{Skumanich} A.,  1972, \mn@doi [\apj] {10.1086/151310}, \href
  {https://ui.adsabs.harvard.edu/\#abs/1972ApJ...171..565S} {171, 565}

\bibitem[\protect\citeauthoryear{{Wahl} et~al.,}{{Wahl}
  et~al.}{2017}]{Wahl2017}
{Wahl} S.~M.,  et~al., 2017, \mn@doi [\grl] {10.1002/2017GL073160}, \href
  {https://ui.adsabs.harvard.edu/abs/2017GeoRL..44.4649W} {44, 4649}

\bibitem[\protect\citeauthoryear{{Weber} \& {Davis}}{{Weber} \&
  {Davis}}{1967}]{Weber1967}
{Weber} E.~J.,  {Davis} Leverett J.,  1967, \mn@doi [\apj] {10.1086/149138},
  \href {https://ui.adsabs.harvard.edu/abs/1967ApJ...148..217W} {148, 217}

\bibitem[\protect\citeauthoryear{{Winn}, {Fabrycky}, {Albrecht}  \&
  {Johnson}}{{Winn} et~al.}{2010}]{Winn2010}
{Winn} J.~N.,  {Fabrycky} D.,  {Albrecht} S.,   {Johnson} J.~A.,  2010, \mn@doi
  [\apjl] {10.1088/2041-8205/718/2/L145}, \href
  {https://ui.adsabs.harvard.edu/abs/2010ApJ...718L.145W} {718, L145}

\bibitem[\protect\citeauthoryear{{Zahn}}{{Zahn}}{2008}]{Zahn2008}
{Zahn} J.~P.,  2008, in {Goupil} M.~J.,  {Zahn} J.~P.,  eds,  EAS Publications
  Series Vol. 29, EAS Publications Series. pp 67--90 (\mn@eprint {arXiv}
  {0807.4870}), \mn@doi{10.1051/eas:0829002}

\makeatother
\end{thebibliography}
\input{main.bbl}


\bsp	
\label{lastpage}
\end{document}